\documentclass{article}
\usepackage{amsmath,amssymb}
\usepackage{graphicx}

\newtheorem{theorem}{Theorem}
\newtheorem{definition}{Definition}
\newtheorem{lemma}[theorem]{Lemma}

\newtheorem{rem}[theorem]{Remark}

\newfont{\bbc}{msbm10}
\newcommand{\inte }{{\rm int}\,}
\newcommand{\dom }{{\rm dom}\,}

\newcommand{\HH}{\mathcal H}

\newcommand{\cover}[1]{\stackrel{#1}{\Longrightarrow}}

\newcommand{\tc}[3]{#1_{#2}^{[#3]}}

\newcommand{\comment}[1]{\mbox{}}

\newcommand{\ltwoq}{l_{2,q}}
\newcommand{\WqS}[2]{W_{#1,#2}}
\newcommand{\Lpow}{p}
\newcommand{\tmax}{t_{\mathrm{max}}}
\newcommand{\llog}{l_{\mathrm{log}}}

\newcommand{\qed}{$\Box$}

\begin{document}
\begin{center}
{\Large \bf Symbolic dynamics for Kuramoto-Sivashinsky PDE on the line --- a
computer-assisted proof}

 \vskip 0.5cm
{\large Daniel Wilczak,  Piotr Zgliczy\'nski \footnote{
Research has been supported by Polish
National Science Centre grant 2016/22/A/ST1/00077.}}

 \vskip 0.2cm
 { Institute of Computer Science and Computational Mathematics,\\
Jagiellonian University}\\
{\small  S. {\L}ojasiewicza 6, 30-348 Krak\'ow, Poland} \\
e-mail: \texttt{\{wilczak,zgliczyn\}@ii.uj.edu.pl, umzglicz@cyf-kr.edu.pl}

\vskip 0.5cm

 \today
\vskip 0.5cm

\end{center}

\begin{abstract}
The Kuramoto-Sivashinsky PDE on the line with  odd and periodic boundary
conditions and with parameter $\nu=0.1212$ is considered. We give a computer-assisted proof the existence
of symbolic dynamics and countable infinity of periodic orbits with arbitrary large periods.
\end{abstract}

\vspace{0.3cm} \noindent {\bf Keywords:}  periodic orbits, dissipative PDEs, Galerkin projection, rigorous numerics, computer-assisted proof

\vspace{0.3cm} \noindent {\bf AMS classification:} 35B40, 35B45,
65G30, 65N30

\vskip\baselineskip

\section{Introduction.}

In the study of nonlinear PDEs, there is a huge gap between  what
we can observe in  numerical simulations and what we can prove
rigorously. One possibility to overcome this problem are computer-assisted
proofs. This paper is an attempt in this direction.

We consider the one-dimensional Kuramoto-Sivashinsky PDE \cite{KT,S} (in the
sequel we will refer to it as the KS equation), which is given by
\begin{equation}
\label{eq:KS} u_t = -\nu u_{xxxx} - u_{xx} + (u^2)_x, \qquad \nu>0,
\end{equation}
where $x \in \mathbb{R}$, $u(t,x) \in \mathbb{R}$ and we impose
odd and periodic boundary conditions
\begin{equation}
  u(t,x)=-u(t,-x), \qquad u(t,x) = u(t,x+ 2\pi). \label{eq:KSbc}
\end{equation}

The Kuramoto-Sivashinsky equation has been introduced by Kuramoto~\cite{KT}  in space dimension one for the study of front propagation in the Belousov-Zhabotinsky reactions. An extension of this equation to space dimension 2 (or more)
has been introduced by G. Sivashinsky \cite{S} in studying the propagation of flame front in the case of mild combustion.

The following theorem is the main result of this paper.

\begin{theorem}\label{thm:main}
The system (\ref{eq:KS}--\ref{eq:KSbc}) with the parameter value $\nu=0.1212$ is chaotic in the following sense.
There exists a compact invariant set $\mathcal A\subset L^2((-\pi,\pi))$ which consists of
\begin{enumerate}
 \item bounded full trajectories visiting explicitly given and disjoint vicinities  of two selected periodic solutions $u^1$ and $u^2$, respectively, with any prescribed order $\{u^1,u^2\}^\mathbb Z$.
 \item countable infinity of periodic orbits with arbitrary large periods. In fact, each periodic sequence of symbols $\{u^1,u^2\}^\mathbb Z$ is realised by a periodic solutions of the system (\ref{eq:KS}--\ref{eq:KSbc}).
\end{enumerate}
\end{theorem}

\begin{figure}[htbp]
\centerline{
\includegraphics[width=.5\textwidth]{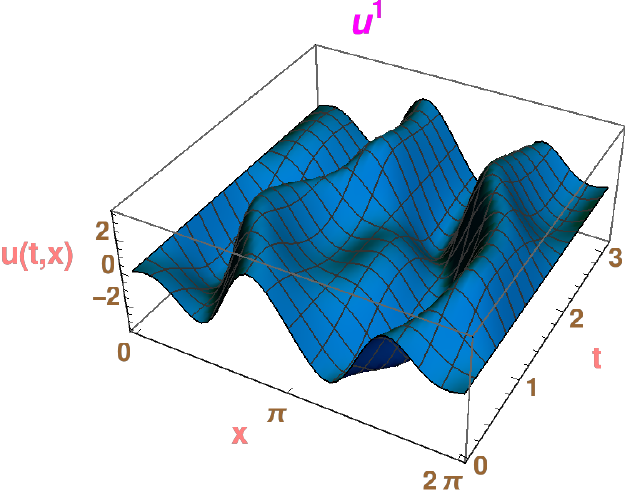}
\includegraphics[width=.5\textwidth]{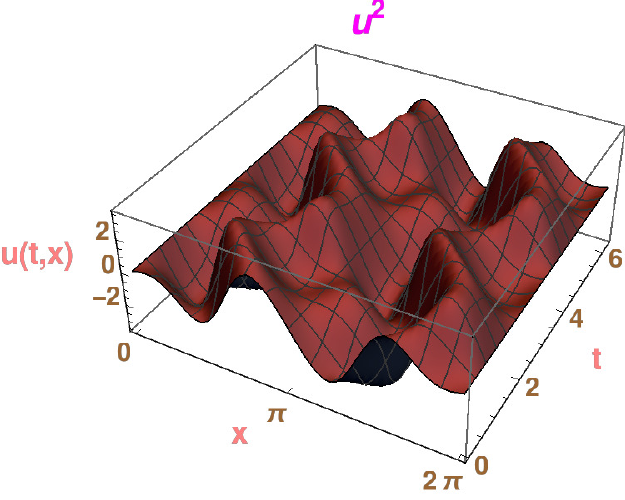}
}
\caption{Two approximate time-periodic orbits $u^1$ and $u^2$.\label{fig:periodic-orbits-3d}}
\end{figure}
\begin{figure}[htbp]
\centerline{\includegraphics[width=.9\textwidth]{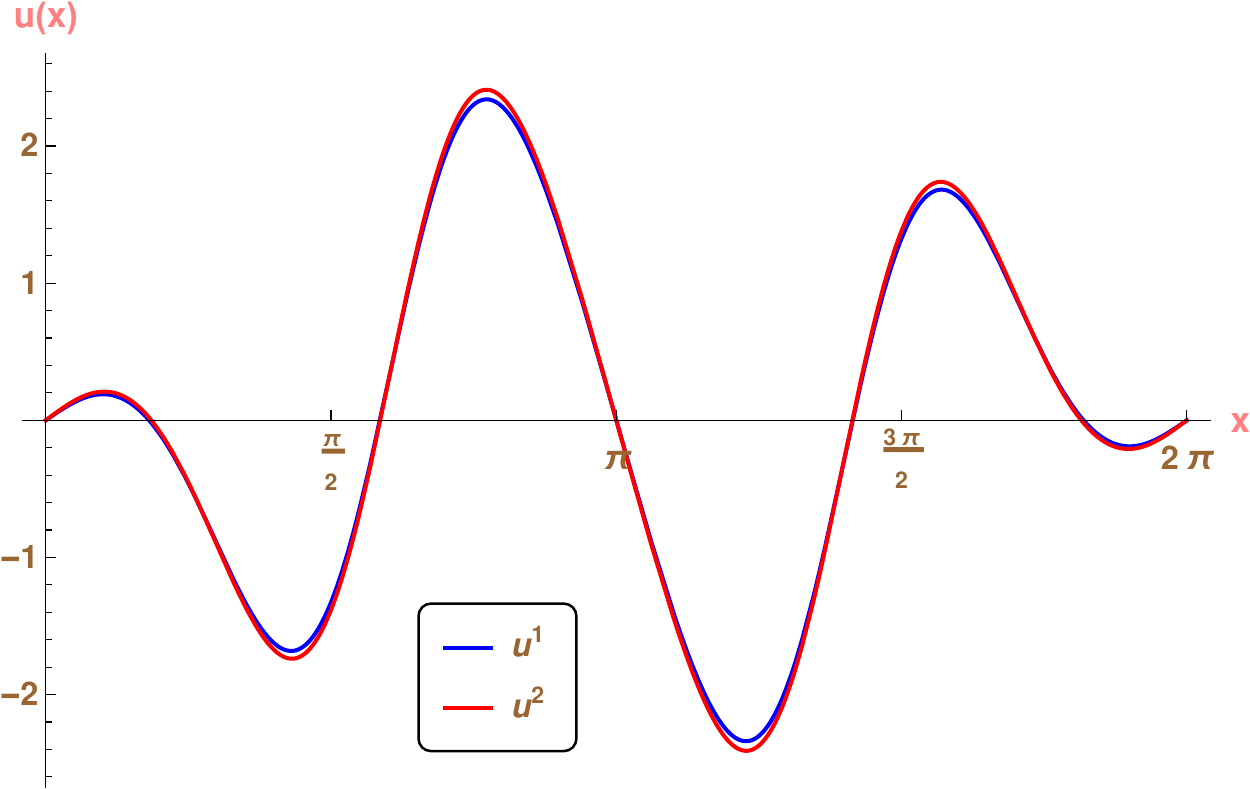}}
\caption{Profiles of initial conditions for time-periodic orbits $u^1$ and $u^2$.\label{fig:periodic-orbits}}
\end{figure}

The two special solutions $u^1$ and $u^2$ appearing in Theorem~\ref{thm:main} are time-periodic
 --- see Fig.\ref{fig:periodic-orbits-3d}.  
Profiles of initial conditions for $u^1$ and $u^2$ 
are shown in Fig.~\ref{fig:periodic-orbits}.

The choice of the KS equation for this study is motivated by the
following facts.
\begin{itemize}
\item The existence of a compact global attractor, the existence of a finite-dimensional inertial manifolds  for (\ref{eq:KS}--\ref{eq:KSbc}) are well
established --- see for example \cite{CEES,FT,FNST,NST} and the
literature cited there. We would like to emphasise, that we are not
using these results in our work.
\item There exist multiple numerical studies of the
dynamics of the KS equation (see for example \cite{CCP,HN,JKT,JJK}), where it
was shown, that the dynamics of the KS equation can be  highly
nontrivial for some values of parameter $\nu$, while being  well
represented by relatively small number of modes.
\item There are  several papers devoted to computer-assisted proofs of periodic
orbits for the KS equation by Zgliczy\'nski \cite{ZKSper,ZKS3}, Arioli and Koch
\cite{AK} and by Figueras, Gameiro, de la Llave and Lessard \cite{FGLL,FL,GaL}.
\end{itemize}

While the choice of the odd periodic boundary conditions was motivated by earlier
numerical studies of KS equation \cite{CCP,JKT},  the basic mathematical
reason is the following: the equation (\ref{eq:KS}) with periodic boundary
conditions has the translational symmetry. This implies, that for
a fixed value of $\nu$, all periodic orbits are members of one-parameter
families of periodic orbits. The restriction to the invariant subspace of odd functions breaks this symmetry, and gives a hope, that dynamically interesting objects are isolated and easier accessible for  computer-assisted proofs.

\begin{figure}[htbp]
\centerline{\includegraphics[width=.9\textwidth]{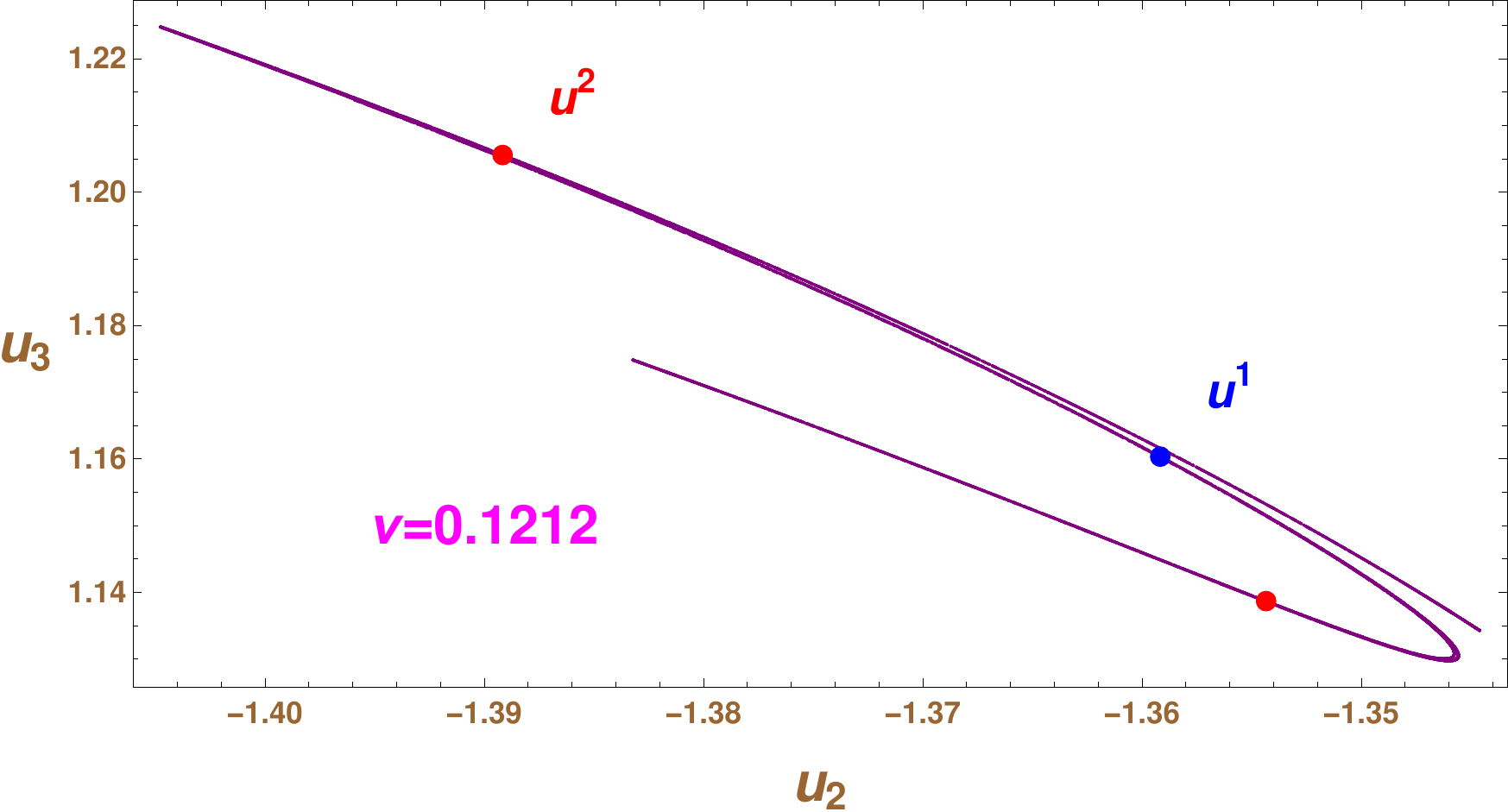}}
\caption{Numerically observed chaotic attractor for (\ref{eq:KS}--\ref{eq:KSbc}) obtained by simulation of a finite-dimensional projection of the corresponding infinite-dimensional ODE for the Fourier coefficients in $u(t,x)=\sum_{k=1}^\infty u_k(t)\sin (kx)$. Projection onto $(u_2,u_3)$ plane of the intersection of the observed attractor with the Poincar\'e section $u_1=0, u_1'>0$ is shown along with an approximate location of the two periodic points $u^1, u^2$ appearing in Theorem~\ref{thm:main}. The point $u^1$ is a fixed point for the Poincar\'e map and $u^2$ is of period two.\label{fig:attractor}}
\end{figure}

Our choice of the parameter value $\nu=0.1212$ is motivated by a numerical  observation, that the Feigenbaum route to chaos through successive period doubling bifurcations \cite{F78}  happens for (\ref{eq:fuKS}) as $\nu$ decreases toward $\nu=0.1212$. For this parameter value a chaotic attractor is observed --- see Fig.~\ref{fig:attractor}.

The proof of Theorem~\ref{thm:main} is a mixture of topological methods and rigorous numerics. It uses neither of the special features of the Kuramoto-Sivashinsky PDE. Therefore, it should be applicable to other systems of dissipative PDEs.

The topological part  exploits the apparent existence of transversal heteroclinic connections between two periodic orbits $u^1$ and $u^2$ in both directions. The two approximate heteroclinic orbits connecting two periodic orbits are then used to obtain a topological horseshoe for some higher iterate of a Poincar\'e map. This is the same type of construction as it is used in the proof of the Smale-Birkhoff homoclinic theorem \cite[Thm.5.3.5]{GH}. There, the existence of symbolic dynamics for a diffeomorphism is concluded from the existence of a transversal intersection of the stable and unstable manifolds of a hyperbolic fixed point.

The numerical part of our approach is an algorithm, which allows to compute rigorous bounds on the trajectories of PDEs with periodic boundary conditions. It is applicable to a class of dissipative PDEs of the following form
\begin{equation}
  u_t = L u + N\left(u,Du,\dots,D^ru\right), \label{eq:genpde}
\end{equation}
where $u \in \mathbb{R}^n$,  $x \in \mathbb{T}^d=\left(\mathbb{R}\mod 2\pi\right)^d$, $L$ is a linear operator, $N$ is a
polynomial and by $D^s u$ we denote $s^{\text{th}}$ order derivative of
$u$, i.e. the collection of all partial derivatives of $u$ of
order $s$.

We require, that the operator $L$ is diagonal in the Fourier basis
$\{e^{ikx}\}_{k \in \mathbb{Z}^d}$,
\begin{equation*}
  L e^{ikx}= -\lambda_k e^{ikx},
\end{equation*}
with
\begin{eqnarray*}
  \lambda_k \approx  |k|^p  \quad\text{and}\quad   p > r.
\end{eqnarray*}

If the solutions are sufficiently smooth, the problem (\ref{eq:genpde}) can be written as an infinite ladder of ordinary differential equations for the Fourier
coefficients in $u(t,x)=\sum_{k \in \mathbb{Z}^d} u_k(t) e^{i kx}$, as follows
\begin{equation}
  \frac{d u_k}{dt} = f_k(u)=-\lambda_k u_k + N_k\left(\{u_j\}_{j \in \mathbb{Z}^d}\right), \qquad \mbox{for all
  $k \in \mathbb{Z}^d$}. \label{eq:fueq}
\end{equation}
The crucial fact, which makes our approach to the rigorous integration of (\ref{eq:fueq})  possible is the \emph{isolation property}, which reads:

\emph{Let}
\begin{equation*}
   W=\left\{ \{u_k\}_{k \in \mathbb{Z}^d}\,|\,  |u_k| \leq \frac{C}{|k|^s q^{|k|}}\right\}
\end{equation*}
\emph{where $q\geq 1$, $C>0$, $s>0$. \\
Then there  exists $K>0$, such that for $|k| > K$ there holds}
\begin{equation*}
 \mbox{if} \quad u \in W, \ |u_k|=\frac{C}{|k|^s q^{|k|}}, \qquad \mbox{then} \qquad  u_k \cdot f_k(u) <0.
\end{equation*}

This property is used in  our algorithm to obtain a priori bounds for $u_k(h)$ for small $h>0$ and $|k|>K$, while the finite number of modes $u_k$ for $|k| \leq K$ is computed using  tools for rigorous integration of ODEs \cite{CAPD,Lo,NJP} based on the interval arithmetics \cite{Mo}.

Our algorithm for rigorous integration of (\ref{eq:fueq}) stems from the  \emph{method of self-consistent  bounds}, which was introduced in \cite{ZM} and later developed in \cite{ZAKS,ZGal,ZNS,ZKSper}. In the present paper we propose some significant modifications to the method, discussed more in details in Section~\ref{sec:algorithm}.

The content of this paper can be described as follows. In Section~\ref{sec:infdim-covrel} we present a new method for proving chaotic dynamics for compact infinite-dimensional maps. In Section~\ref{sec:application} we give an application of this method to a family of Poincar\'e maps for an infinite-dimensional ODE associated to (\ref{eq:KS}--\ref{eq:KSbc}) and we give a proof of Theorem~\ref{thm:main}.
In the remaining sections we outline the algorithm for rigorous integration forward in time of dissipative PDEs on the example of the system (\ref{eq:KS}--\ref{eq:KSbc}).

\subsection{Notation.}

By $\mathbb{C}$, $\mathbb{R}$, $\mathbb{Z}$, $\mathbb{N}$ we denote the sets of
complex, real, integer and natural numbers including zero, respectively. With $\mathbb{N}_+$ we denote positive integers.

Let $(T,\rho)$ be a metric space. For a set $X \subset T$, by $\inte X$,
$\overline{X}$ and $\partial X$ we denote the interior, the closure and the
boundary of $X$, respectively. If $X \subset Y \subset T$, then by
$\inte_Y X$ and by $\partial_Y X$ we denote respectively the
interior and the boundary of $X$ with respect to the metric space
$(Y,\rho)$. By $B(c,r)=\{ x\, |\, \rho(c,x) < r \}$ we denote
the ball of radius $r$ centred at $c$. For a point $p \in T$ put
$\rho(p,X)=\inf\{ \rho(p,q)\, |\,  q \in X \}$. We
 define $B(X,\epsilon)=\{ y\, |\, \rho(y,X) < \epsilon  \}$.

For $k \in \mathbb{N}$, $c\in \mathbb{R}^k$ and $r \geq 0$ by $B_k(c,r)$ we denote an open
ball in $\mathbb{R}^k$ of the radius $r$ and the centre $c$. The norm used to
define $B_k(c,r)$ will be usually the euclidean one, but in most cases in this paper any norm can be used. By $B_k=B_k(0,1)$ and $\overline{B_k}=\overline{B_k}(0,1)$ we will denote the open and closed unit balls, respectively.


Let $\mathcal H=\mathcal X\oplus \mathcal Y$ be a direct sum of two orthogonal subspaces
$\mathcal X,\mathcal Y\subset \HH$. We extend the notion of the direct sum to
sets in a natural way. For $A\subset \mathcal X$ and $B\subset \mathcal Y$ we
set
$$A\oplus B := \{a+b \,|\, a\in A, b\in B \}.$$

We will be often dealing with Galerkin projections in some space $\mathcal H$ with the basis $\{e_i\}_{i \in J}$, $J\subset \mathbb{Z}^d$.  We define the following projections:
$\pi_k \left(\sum_{i\in J}u_i e_i\right)=u_k $ and $\pi_{\leq n} \left(\sum_{i\in J}u_i e_i\right)= \sum_{|i| \leq n}u_i e_i $ and analogously for other possible sets of indices  entering into the projection. For a set $W \subset \pi_{\leq n} X$ by $\inte_{\leq n} W$ and $\partial_{\leq n}W$ we will denote respectively the interior and the boundary of $W$ with respect to the set $\pi_{\leq n} X$.

\section{Topological method for symbolic dynamics for maps in infinite dimension.}
\label{sec:infdim-covrel}
Topological methods proved to be very useful in the context of computer-assisted study of dynamical systems. One of the most efficient in the context of studying chaotic dynamics is the \emph{method of covering relations} introduced in \cite{Z0,Z4} for maps with one exit ("unstable") direction and later extended to include many exit directions in \cite{ZGi}  known also in the literature  as the method of correctly aligned windows \cite{E1,E2}. In this section we extend this method to a class of maps defined on compact (possible infinite-dimensional) subsets of real normed spaces.  The notion is quite similar but relies on the fixed point index  maps on compact  ANRs as defined in the book of Granas and  Dugundji \cite{GD} (see also survey by Mawhin \cite{M}). Such an extension is motivated by the applications we keep in mind -- Poincar\'e maps for infinite-dimensional ODEs. Another variant of this extension was discussed in  \cite{Z1,ZKS3}.

\subsection{Covering relations in compact ANRs.}

The aim of this section is to extend the notion of covering relation to infinite-dimensional case.
\begin{definition}\label{def:hset}
  Let $X$ be a real normed space. A $h$-set, $N=(|N|,c_N,u(N))$, is an object consisting of the following data
\begin{itemize}
 \item $|N|\subset X$ -- a compact set called \emph{the support} of $N,$
 \item $u(N)$ -- a nonnegative integer,
 \item $c_N$ -- a homeomorphism of $X$ such that
 \begin{equation*}
  c_N^{-1}(|N|) = \overline{B_{u(N)}}\oplus T_N =: N_c,
 \end{equation*}
 where $T_N$ is a convex set.
\end{itemize}
\end{definition}
The above definition generalizes the concept of $h$-sets introduced in \cite{ZGi} for finite-dimensional spaces, where a $h$-set is defined as the product of closed unit balls $\overline{B}_u\times \overline{B}_s\subset \mathbb R^{u+s}$ in a coordinate system $c_N$.  The next definition extends the notion of covering relations from \cite{ZGi} to compact maps acting on infinite-dimensional real normed spaces.

\begin{definition}\label{def:covrel}
  Let $N$ and $M$ be h-sets in $X$ and $Y$, respectively such that $u=u(N)=u(M)$. Let $f\colon |N|\to Y$ be a continuous map and set $f_c:=c_M^{-1} \circ f\circ c_N$. We say that $N$ $f$-covers $M$, denoted by $N\cover{f}M$,  if there is a linear map $L:\mathbb R^u\to \mathbb R^u$ and a compact homotopy $H:[0,1]\times N_c\to Y$, such that
\begin{quote}
\begin{itemize}
 \item[{\rm\bf[CR1]:}] $H(0,\cdot) = f_c$,
 \item[{\rm\bf[CR2]:}] $H(1,x,y) = (L(x),0)$, for all $(x,y)\in N_c$,
 \item[{\rm\bf[CR3]:}] $H(t,x,y) \notin M_c$, for all $(x,y)\in \partial{B_u}\oplus T_N$, $t\in[0,1]$ and
 \item[{\rm\bf[CR4]:}] $H(t,x,y) \in \mathbb{R}_u\oplus T_M$ for $(x,y)\in N_c$ and $t\in[0,1]$.
\end{itemize}
\end{quote}
\end{definition}

\begin{figure}[hptb]
\centerline{\includegraphics[width=\textwidth]{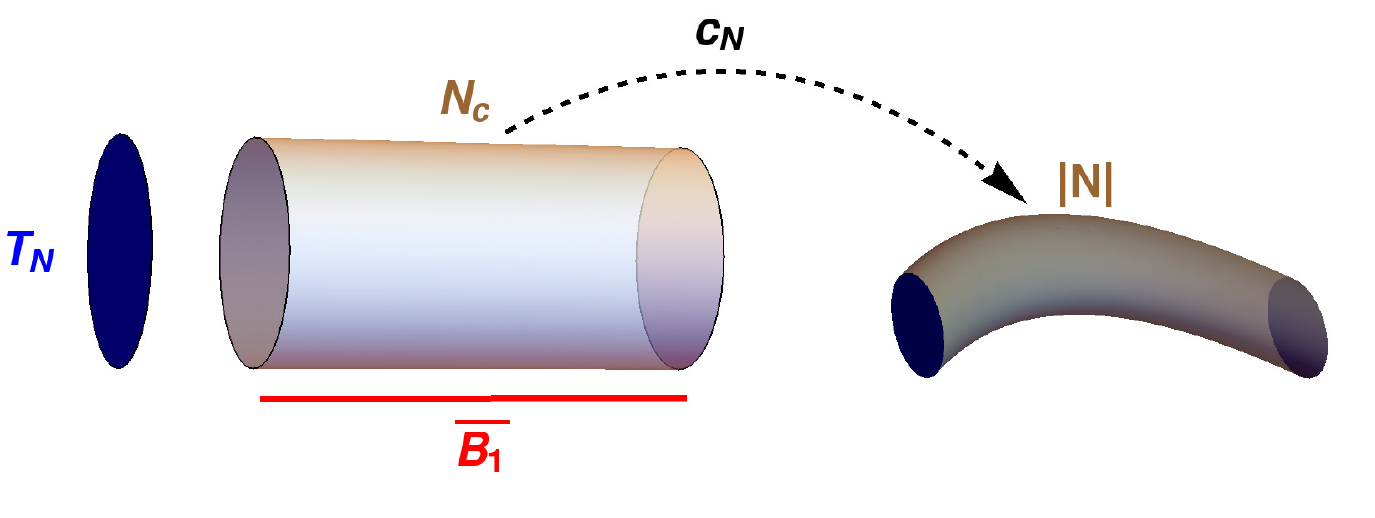}}
\caption{An example of an h-set in three dimensions with $u(N)=1$ and $T_N=D_2$ -- a two-dimensional closed disc. Here $N_c=\overline{B_1}\oplus D_2$.}
\label{pic:magicset}
\end{figure}

\begin{figure}[hptb]
\centerline{\includegraphics[width=.9\textwidth]{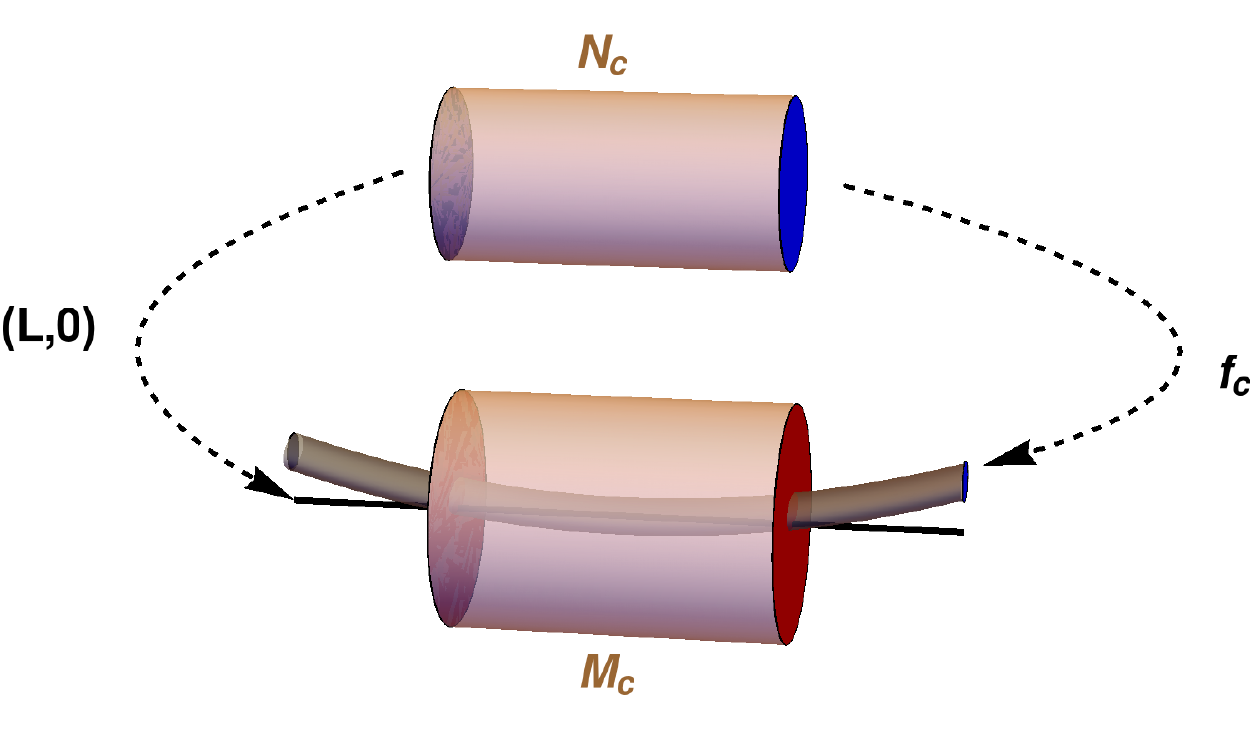}}
\caption{An example of an $f-$covering relation: $N\cover{f}M$. In this case, the homotopy joining $f_c(x,y)$ with a linear map $(L(x),0)$ and satisfying \textbf{[CR1]}--\textbf{[CR4]} is simply given by $H(t,x,y) = t(L(x),0) + (1-t)f_c(x,y)$.}
 \label{fig:cov}
\end{figure}

A typical picture of a h-set with $u(N)=1$ is given in Figure \ref{pic:magicset}. A  picture illustrating covering relation with one exit direction is given on Figure~\ref{fig:cov}.

The following theorem extends applicability of the notion of covering relation to compact maps in real normed spaces.
\begin{theorem}\label{thm:periodic-covrel}
  Let $X_i$, $i=1,\ldots,k$ be real normed spaces and let $N_i$ be $h$-sets in $X_i$, respectively. Assume
 \begin{equation*}
  N_1\cover{f_1}N_2\cover{f_2}\cdots\cover{f_{k-1}}N_{k}\cover{f_k}N_1.
 \end{equation*}
Then there exists $u_1\in |N_1|$ such that
\begin{eqnarray}
 (f_i\circ \cdots \circ f_1)(u_1)&\in& |N_{i+1}|\quad \text{for } i=2,\ldots,k-1\quad \text{and}\label{eq:po1}\\
 (f_k\circ \cdots \circ f_1)(u_1)&=&u_1\label{eq:po2}.
\end{eqnarray}
\end{theorem}
\textbf{Proof:}
Put $N_{k+1}=N_1$. The set $X=X_1\times \ldots\times  X_k$ with the maximum norm is a real normed space. The set
$$Y=\left(\mathbb R^u\oplus T_{N_1}\right)\times\cdots \left(\mathbb R^u\oplus T_{N_k}\right)\subset X$$
is convex and by \cite[Col.~4.4]{GD} it is an ANR. Put $$N=\left(B_u\oplus T_{N_1}\right)\times\cdots\times \left (B_u\oplus T_{N_k}\right)\subset Y.$$

Let $H_i$ be a homotopy from the definition of covering relation $N_i\cover{f_i}N_{i+1}$.
By \textbf{[CR4]} the range of $(f_i)_c$ and $H_i(t,\cdot)$ is in $\mathbb R^u\oplus T_{N_{i+1}}$ for $i=1,\ldots,k$, $t\in[0,1]$. Therefore we can define a map $F\colon \overline{N}\to Y$ by
$$
F\left(u_1, u_2,\ldots,u_k\right) = \left((f_{k})_c(u_k),(f_1)_c(u_1),\ldots ,(f_{k-1})_c(u_{k-1})\right)
$$
and a homotopy
$H\colon[0,1]\times \overline{N}\to Y$ by

$$
H\left(t,\begin{bmatrix}
(x_1,y_1)\\ (x_2,y_2)\\ \ldots\\(x_k,y_k)
\end{bmatrix}\right) =
\begin{bmatrix}
H_{k}(t,x_k,y_k)\\
H_1(t,x_1,y_1)\\
\ldots \\
H_{k-1}(t,x_{k-1},y_{k-1})
\end{bmatrix}.
$$
It is easy to see that for all $t\in[0,1]$ the mapping $H(t,\cdot)$ is fixed point free on
 $\partial_Y N$. Indeed, if $u=\left((x_1,y_1), \ldots\\(x_k,y_k)\right)\in \partial_Y\overline{N}$ then $x_i\in\partial B_u$ for some $i=1,\ldots,k$. From \textbf{[CR3]}
 we have $H_i(t,x_i,y_i)\notin \overline{B_u}\oplus T_{N_{i+1}}$ and therefore $H(t,u)\notin \overline{N}$.

Thus, the fixed point index $i(H(t,\cdot),N)$ for maps on compact ANRs \cite{GD} is well defined and does not depend on $t\in[0,1]$.

For $t=1$ we have
$$
H\left(1,\begin{bmatrix}
(x_1,y_1)\\\ldots\\(x_k,y_k)
\end{bmatrix}\right) =
\begin{bmatrix}
(L_k(x_k),0)\\
(L_1(x_1),0)\\
\ldots \\
(L_{k-1}(x_{k-1}),0)
\end{bmatrix},
$$
where $L_i$ is a linear map from the covering relation $N_i\cover{f_i} N_{i+1}$. By \textbf{[CR2]} it is an expanding isomorphism which maps $\partial B_u$ out of the unit ball $\overline{B_u}$.
Thus, $H(1,\cdot)$ is a hyperbolic linear map and
$$
\left|i\left(H(1,\cdot),N\right)\right| = 1.
$$
Therefore $i\left(F,N\right)=I\left(H(0,\cdot),N\right)\neq 0$ and in consequence the mapping $F$ has a fixed point $(\hat u_1,\ldots,\hat u_k)\in N$. Now the point $u_1=c_{N_1}(\hat u_1)$ satisfies
(\ref{eq:po1}) and (\ref{eq:po2}).
{\flushright{\qed}}

\subsection{Subshifts of finite type and symbolic dynamics.}
The following definitions are standard (see for example
\cite{GH}). Let us fix $k>0$ and let
$\left(A_{ij}\right)_{i,j=1,\dots,k}$ be $k \times k$ matrix, such that
$A_{ij} \in \{0,1\}$. We define $\Sigma_A$ and $\Sigma_A^+$ by
\begin{eqnarray*}
  \Sigma_A &=& \{ c \in \{1,2,\dots,k\}^\mathbb{Z} \ | \ A_{c_i c_{i+1}}=1 \
\forall i \in \mathbb{Z}
  \}, \\
  \Sigma_A^+ &=& \{ c \in \{1,2,\dots,k\}^\mathbb{N} \ | \ A_{c_i c_{i+1}}=1
\ \forall i \in \mathbb{N}
  \}.
\end{eqnarray*}
We define a shift map $\sigma$ on $\Sigma_A$ and $\Sigma_A^+$ by
\begin{equation*}
  \sigma(c)_i= c_{i+1}, \quad \forall i \in \mathbb{Z} \ (i \in \mathbb{N}).
\end{equation*}
The pairs $(\Sigma_A,\sigma)$ and $(\Sigma^+_A,\sigma)$ are called
\emph{subshifts of finite type with transition matrix $A$}.  Let
\begin{eqnarray*}
  \Sigma_k &=& \{1,2,\dots,k\}^\mathbb{Z},\\
  \Sigma_k^+ &=& \{1,2,\dots,k\}^\mathbb{N}.
\end{eqnarray*}
We call $(\Sigma_k,\sigma)$ and $(\Sigma_k^+,\sigma)$ \emph{full
shifts} on $k$ symbols.

Let $X_i$, $i=1,\ldots,k$ be real normed spaces and let $N_i\subset X_i$ be $h$-sets, such that  $|N_i| \cap |N_j| = \emptyset$ for $i\neq j$. Note, we do not require that the spaces $X_i$ are different. Let $f_i:|M_i|\to Y_i$ be continuous maps, for $i=1,\ldots, m$, where $M_i\in\{N_1,\ldots,N_k\}$ and $Y_i\in\{X_1,\ldots,X_k\}$. Again, we accept that two different mappings are defined on the same set $|N_i|$.

We define a \emph{transition matrix} $\left(A_{ij}\right)_{i,j=1,\ldots,k}$ in the following way:
\begin{equation*}
 A_{ij} = \begin{cases}
  1 & \text{if } N_i\cover{f_c} N_j, \text{ for some } c\in\{1,\ldots,m\},\\
  0 & \text{otherwise}.
 \end{cases}
\end{equation*}

\begin{definition}
 A sequence $(x_i)_{i\in\mathbb N}$ is called a full trajectory with respect to family of maps $(f_1,\ldots,f_m)$ if for all $i\in\mathbb N$ there is $c\in\{1,\ldots,m\}$ such that
 $$
 f_c(x_i) = x_{i+1}.
 $$
 A sequence $(\alpha_i)_{i\in\mathbb N}\in\{1,\ldots,k\}^{\mathbb N}$ is called admissible, if there is a full trajectory $(x_i)_{i\in\mathbb N}$ with respect to $(f_1,\ldots,f_m)$ such that
 $$x_i\in |N_{\alpha_i}|.$$
\end{definition}

The following theorem address the issue of the existence of an orbit realising a non-periodic sequence of covering relations.
\begin{theorem} \label{thm:symbolic-dynamics}
Every sequence of symbols
$(\alpha_i)_{i\in\mathbb N}\in \Sigma_A^+$ is admissible. Moreover, if $(\alpha_i)_{i\in\mathbb N}$ is $T$-periodic, then the corresponding trajectory $(x_i)_{i\in\mathbb N}$ may be chosen to be a $T$-periodic sequence, too.
\end{theorem}
\textbf{Proof:}
From Theorem~\ref{thm:periodic-covrel} every $T$-periodic sequence $(\alpha_i)_{i\in\mathbb N}\in \Sigma_A^+$ of symbols is realized by a $T$-periodic trajectory $(x_i)_{i\in\mathbb N}$.

Let us fix a sequence $(\alpha_i)_{i\in\mathbb N}\in \Sigma_A^+$ which is not periodic. This means, that there is a sequence $(c_i)_{i\in\mathbb N}$ such that
$$N_{\alpha_i}\cover{f_{c_i}}N_{\alpha_{i+1}}$$
for $i\in\mathbb N$. For every $T>0$ we can construct a closed loop of covering relations
$$N_{\alpha_0}\cover{f_{c_0}}N_{\alpha_1}\cover{f_{c_1}}\ldots \cover{f_{c_{T-1}}}N_{\alpha_T}\cover{g}N_{\alpha_0},$$
by adding an artificial covering relation (an affine map $g$) at the end of this sequence. From Theorem~\ref{thm:periodic-covrel} this sequence is realised by a $(T+1)$-periodic full trajectory $(x_i^T)_{i\in \mathbb N}$. Since $|N_{\alpha_0}|$ is compact, we can choose a subsequence  $(x_0^{T_j})_{j\in\mathbb N}$ which converges to $x_0\in |N_{\alpha_0}|$. Then for $i\in\mathbb N$ and $T_j> i$ we have $(f_{c_i}\circ\ldots \circ f_{c_0})(x_0^{T_j})\in |N_{\alpha_{i+1}}|$ and by the continuity of each $f_{c_i}$ we obtain that
\begin{eqnarray*}
x_{i+1}&:=&(f_{c_i}\circ\ldots\circ f_{c_0})(x_0)=\lim_{j\to\infty }(f_{c_i}\circ\ldots \circ f_{c_0})(x_0^{T_j})\in |N_{\alpha_{i+1}}|.
\end{eqnarray*}
The constructed sequence $(x_i)_{i\in\mathbb N}$ satisfies the assertion.
\qed

\section{The KS equation.}\label{sec:application}

Consider the equation (\ref{eq:KS}) with periodic and odd boundary conditions (\ref{eq:KSbc}) and assume that $u$ is a solution to (\ref{eq:KS}--\ref{eq:KSbc}) given as a convergent Fourier series
\begin{equation}
u(t,x)=\sum_{k=1}^\infty -2a_k(t) \sin(kx).  \label{eq:ks-func-rep}
\end{equation}

The particular form of representation of $u$ given in (\ref{eq:ks-func-rep})  comes from imposing on $u(t,x)=\sum_k u_k(t) e^{ikx}$ periodic and odd boundary conditions. Then we have  $a_k(t)=\mathrm{Im}\, u_k(t) $.

It is easy to see that (see for example \cite{CCP,ZM}) that  for sufficiently regular functions  the system (\ref{eq:KS}--\ref{eq:KSbc})  give rise  to an infinite ladder of coupled ODEs 
\begin{equation} \label{eq:fuKS}
  \frac{d a_k}{dt}=k^2(1-\nu k^2) a_k - k \sum_{n=1}^{k-1} a_n
  a_{k-n} + 2k \sum_{n=1}^{\infty} a_n  a_{n+k}, \quad k=1,2,3\dots
\end{equation}

In order to apply topological fixed point theorems discussed in Section~\ref{sec:infdim-covrel} to (\ref{eq:fuKS}) we need to chose  a topology for space of sequences $\left(a_k\right)_{k=1}^\infty$. We will demand that $\left(a_k\right)_{k=1}^\infty \in l_2$, where
\begin{eqnarray*}
 l_2 &=& \left\{(a_k)_{k=1}^\infty\, |\, \sum_{k=1}^\infty|a_k|^2<\infty\right\}, \\
\|(a_k)_{k=1}^\infty\|_2 &=& \sqrt{\sum_{k=1}^\infty|a_k|^2}.  
\end{eqnarray*}
The use of $l_2$ defined in terms of $(a_k)_{k=1}^\infty$ is quite arbitrary. More natural function spaces and norms (compare Theorem~\ref{thm:ks-semi-dyn-pde} in the Appendix~\ref{subapp:KS-existence}) would be the ones induced from $L^2$ or $H^k$ with $k\geq 1$ on the function space  of $2\pi$-periodic functions. This, however, does not matter much, as we will consider the geometrically fast decaying sequences in $l_2$, only. Geometric decay of coefficients guarantees that these sequences represent analytic functions for which replacing (\ref{eq:KS}) with (\ref{eq:fuKS}) makes sense.

For $S>0$ and $q>1$ we set
\begin{eqnarray*}
 \WqS{q}{S} &=& \left\{(a_k)_{k=1}^\infty \,|\, a_k \in \mathbb{R}, \ |a_k|\leq Sq^{-k}\text{ for }k\geq 1\right\},\\
 \ltwoq &=& \left\{(a_k)_{k=1}^\infty \,|\, a_k \in \mathbb{R}, \  \exists S\geq 0 : |a_k|\leq Sq^{-k}\text{ for }k\geq 1\right\} = \bigcup_{S>0}\WqS{q}{S}.
\end{eqnarray*}
Observe the set $\WqS{q}{S} \subset l_2$ is compact. It is also compact   in the topology induced by the $H^k$ norm or $L^2$ norm on the space $2\pi$-periodic functions and the convergence on $\WqS{q}{S}$ in any of these norms and in $l_2$ norm is equivalent to the coordinate-wise convergence, i.e. $\lim_{j\to\infty}\left(a_k^{j}\right)_{k\in\mathbb N_+} = \left(c_k\right)_{k\in\mathbb N_+}$ iff for all $k \in \mathbb{N}_+$ holds $\lim_{j\to\infty}a_k^{j} = c_k$.

From classical results (see Theorem~\ref{thm:ks-semi-dyn-pde} in Appendix~\ref{subapp:KS-existence}) it follows that forward solutions of system (\ref{eq:fuKS})  exist for all initial conditions $a \in l_2$ which defines a continuous semiflow on $l_2$ (a semigroup in the terminology used in \cite{Temam}).
In fact, we are not using this result in the sequel. For our proof we use the local semiflow restricted to some  $\WqS{q}{S}$ for suitably chosen $S$ and $q$.

\subsection{Symbolic dynamics in the KS equation.}

\subsubsection{Heuristics for Galerkin projection of (\ref{eq:fuKS}).}
\label{sec:heuristics}
\begin{figure}[htbp]
\centerline{\includegraphics[width=\textwidth]{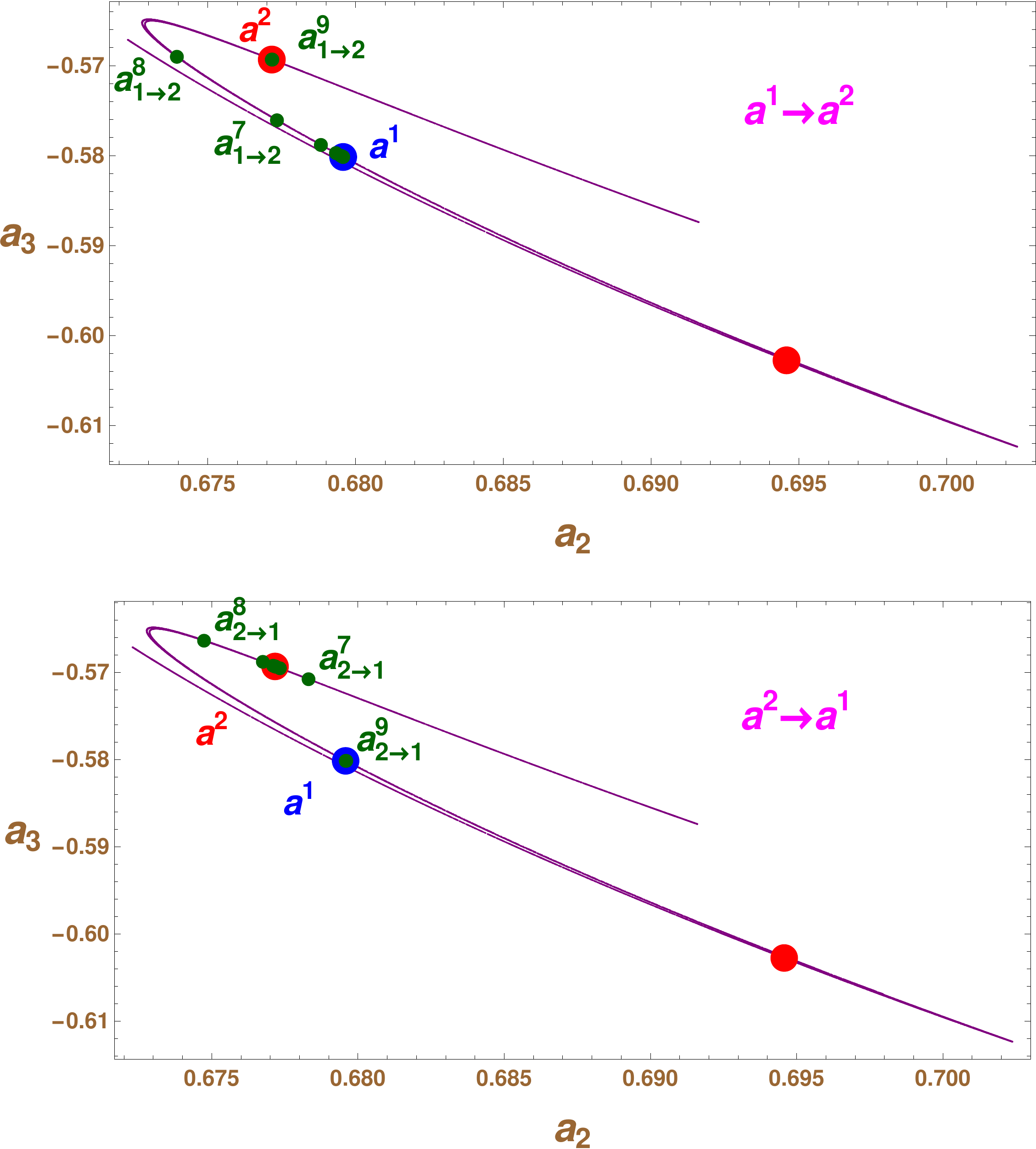}}
\caption{Numerically observed heteroclinic connections for $P$ between $a^1$ and $a^2$ in both directions.\label{fig:heteroclinic}}
\end{figure}

After an extensive numerical simulation we have found that the dynamics of Galerkin projections does not differ significantly for projections of dimension $n\geq 14$. Therefore, we set $n=14$ and we constructed all objects that appear in our computation (like Poincar\'e sections, h-sets) using approximate flow generated by $n$-dimensional projection.

Let us consider a Poincar\'e section
$$\Theta=\{(a_k)_{k=1}^\infty\in \pi_{\leq n}l_2\,|\,a_1=0\,\wedge\,a_1'>0\}$$
and the associated Poincar\'e map $P:\Theta \supset \dom(P) \to\Theta$ with respect to the  local flow induced by the $n$-dimensional Galerkin projection.

After an extensive numerical simulation we have found approximate periodic and heteroclinic points for $P$ -- see Fig.~\ref{fig:heteroclinic} and supplementary material \cite{W}. These are
\begin{itemize}
\item $a^1 \in\Theta$ -- an approximate fixed point for $P$, with one unstable direction,
\item $a^2 \in\Theta$ -- an approximate period-two point for $P$, with one unstable direction,
\item $a_{1\to2}^i\in\Theta$, $i=0,\ldots,10$ -- an approximate heteroclinic chain satisfying
\begin{eqnarray*}
\|a^1-a_{1\to2}^0\|_2&\approx&1.676\cdot 10^{-6},\\
P^2(a_{1\to2}^i)&\approx& a_{1\to2}^{i+1}\quad\text{for}\quad i=0,\ldots,9,\\
\|P^2(a_{1\to2}^{10})- a^2\|_2&\approx&3.932\cdot10^{-10},
\end{eqnarray*}
\item $a_{2\to1}^i\in\Theta$, $i=0,\ldots,10$ -- an approximate heteroclinic chain satisfying
\begin{eqnarray*}
\|a^2-a_{2\to1}^0\|_2&\approx&2.425\cdot10^{-6},\\
P^2(a_{2\to1}^i)&\approx& a_{2\to1}^{i+1}\quad\text{for}\quad i=0,\ldots,7\text{ and } i=9,\\
P^3(a_{2\to1}^8)&\approx& a_{2\to1}^9,\\
\|P(a_{2\to1}^{10})- a^1\|_2&\approx&1.144\cdot10^{-6}.
\end{eqnarray*}
\end{itemize}

\subsubsection{Full system (\ref{eq:fuKS}).}

We consider the system (\ref{eq:fuKS}) on $\ltwoq$ with $q=3/2$. In what follows we will use approximate periodic points $a^1, a^2$ for $P$ and heteroclinic chains $a_{1\to 2}^i, a_{2\to 1}^i$, $i=0,\ldots,10$  as presented in Section~\ref{sec:heuristics}.

Analogously as in the case ODEs we can define the Poincar\'e map between two sections, see Appendix~\ref{subapp:poincarePDE}. The notation is the same as for ODEs. We define
$$P_{\Pi_1\to\Pi_2}^i= P^{i-1}_{\Pi_2 \to \Pi_2}  \circ   P_{\Pi_1 \to \Pi_2}.  $$

The approximate heteroclinic loop for $n$-dimensional Galerkin projection was used to construct symbolic dynamics for the infinite-dimensional system (\ref{eq:fuKS}). More precisely, we constructed
\begin{itemize}
 \item affine Poincar\'e sections $\Theta^1$ and $\Theta^2$, such that $a^i\in\Theta^i$, $i=1,2$,
 \item affine Poincar\'e sections $\Theta_{1\to2}^i$ and $\Theta_{2\to1}^i$,  such that $a_{1\to 2}^i\in\Theta_{1\to 2}^i$ and $a_{2\to 1}^i\in\Theta_{2\to 1}^i$ for $i=0,\ldots,10$,
 \item two $h$-sets $N^i\subset \Theta^i$, $i=1,2$ with $u(N^i)=1$ (i.e. one exit direction), such that $a^i\in |N^i|$, respectively,
 \item two sequences of $h$-sets $N_{1\to2}^i\subset \Theta_{1\to2}^i$ and $N_{2\to1}^i\subset \Theta_{2\to1}^i$ for $i=0,\ldots,10$, with $u(N_{1\to2}^i)=u(N_{2\to1}^i)=1$, such that $a^i_{j\to c}\in |N^i_{j\to c}|$.
\end{itemize}
Explicit coordinates of the points $a^i, a^i_{j\to c}$, the Poincar\'e sections $\Theta^i, \Theta^i_{j\to c}$ and all $h$-sets listed above are given in the supplementary material \cite{W}.

\begin{rem}
 One important parameter in our computer-assisted proof is an integer $m$ which is by our choice set to $m=n+1=15$. This is the number of explicitly stored Fourier coefficients of $(a_k)_{k=1}^\infty$ in a computer memory. The remaining coefficients, called tail, are bounded uniformly by a set represented by two real numbers $S$ and $q$ --- see representation of $h$-sets in Appendix \ref{subapp:covrel-check}. 
\end{rem}

\begin{rem}
The Poincar\'e section which contains a point $p\in\{a^1,a^2,a_{1\to2}^i,a_{2\to1}^i\}$ is chosen to be a hyperplane almost orthogonal to the flow direction of $m$-dimensional Galerkin projection of (\ref{eq:fuKS}) at the point $p$. This means, in particular, that all Poincar\'e sections are defined in terms of $\left(a_k\right)_{k=1}^{m}$ by
$$\left\{x \in \ltwoq\, |\, \sum_{k=1}^{m} f_k(p)(x - p)_k=0\right\}.$$

We have found, that locally orthogonal sections help us in reducing overestimation when we compute rigorously Poincar\'e map.
\end{rem}
\begin{rem}
All $h$-sets $N^i$, $N_{j\to c}^i$ are constructed to be pairwise disjoint.
\end{rem}

Using an algorithm for rigorous integration of (\ref{eq:fuKS}) discussed in the next sections we obtained a computer-assisted proof of the following lemma.
\begin{lemma}\label{lem:symbolic-dynamics}
All the covering relations listed below are satisfied.
\begin{multline*}
  N^1\cover{P^2}N^1 \cover{P^3} N^0_{1\to 2} \cover{P^5} N^1_{1\to 2}\cover{P^5}N^2_{1\to 2}\cover{P^5}\cdots \cover{P^5} N^{10}_{1\to 2}\cover{P^5} N^2,\\
  N^2\cover{P^4}N^2 \cover{P^4} N^0_{2\to 1} \cover{P^5} N^1_{2\to 1}\cover{P^4}N^2_{2\to 1}\cover{P^5}\cdots \\
  \cdots \cover{P^4} N^{8}_{2\to 1}\cover{P^6} N^{9}_{2\to 1}\cover{P^5} N^{10}_{2\to 1}\cover{P^3} N^1,
\end{multline*}
where the starting and target sections in each of the above covering relations are determined by the h-sets appearing in the relation.
\end{lemma}
The conditions to check the covering relation with one exit direction are discussed in Appendix~\ref{subapp:covrel-check}. Some details and technical data regarding computer-assisted verification of Lemma~\ref{lem:symbolic-dynamics} are given in Appendix \ref{sec:technical-data} and the supplementary material \cite{W}.

The subsequent remarks explain the choices made in construction of sequences of covering relations in Lemma~\ref{lem:symbolic-dynamics}.

\begin{rem}
 The largest (in absolute value) eigenvalues of $DP(a^1)$ and $DP^2(a^2)$ are approximately $-1.77$ and $-2.57$, respectively. This suggests, that the expansion of $P$ in the domain we are interested in is rather weak. The second iterate squares both expansion and contraction factors. In consequence, stronger hyperbolicity lets the dynamics to help us in rigorous validation of covering relations for $P^2$ or higher iterates --- there are  wider margins for unavoidable overestimation when compute rigorously the image of an h-set. Another consequence of choosing $P^2$ is the reduction of the number of h-sets along heteroclinic chains that we have to construct.
\end{rem}

\begin{rem}
The particular choice of the third iteration when computing $P^3(a_{2\to1}^8)$ comes from the fact, that it was much easier to construct coordinate system for an $h$-set centred at the point $P^3(a_{2\to1}^8)$ than at $P^2(a_{2\to1}^8)$. This is due to the fact, that $P^2(a_{2\to1}^8)$ is close to the bend in the attractor, while $a^9_{2\to1}$ is already very close to the fixed point $a^1$.
\end{rem}

Observe that in the sequence of covering relations in Lemma~\ref{lem:symbolic-dynamics} there appear various iterates of $P$, while from the construction of the centres of $N^i$ and $N^i_{j\to c}$ one would expect that there  should be just the coverings for $P$, $P^2$ and $P^3$. The reasons are as follows.

First, in the definitions of sections $\Theta^i$, $\Theta^i_{1 \to 2}$ and $\Theta^i_{2 \to 1}$ we do not specify the \emph{crossing direction} -- intersections in both directions are possible. Thus, fixed point for $P$ becomes period-two point for $P_{\Theta^1\to\Theta^1}$ and similarly $a^2$ is period-four point for $P_{\Theta^2\to\Theta^2}$.

Another reason is that in (\ref{lem:symbolic-dynamics}) the maps $P^j$ are not iterates of the Poincar\'e map on the section $\Theta=\{a_1=0, a_1'>0\}$, but they are compositions of $j$ Poincar\'e maps between sections  $\Theta^i$, $\Theta^i_{1 \to 2}$ or $\Theta^i_{2 \to 1}$. Some of these sections are aligned so that they are almost parallel and close. For example, in $N^1\cover{P^3}N_{1\to2}^0$, a trajectory starting from $N^1$ almost immediately cuts $\Theta_{1\to 2}^0$ but this is not what we wanted. Then we follow the trajectory until it intersects twice $\Theta_{1\to2}^0$ in the vicinity of $a^0_{1\to2}$ --- note, we allow intersections with section in both directions.

\textbf{Proof of Theorem~\ref{thm:main}:} Apply  Theorem~\ref{thm:symbolic-dynamics} to the transition matrix coding the covering relations
from Lemma~\ref{lem:symbolic-dynamics}.

In particular, the existence of the "selected" periodic orbits $u^1$ an $u^2$ follows from the coverings $ N^1\cover{P^2}N^1$ and $ N^2\cover{P^4}N^2$.
\qed

\section{The algorithm for rigorous bounds for solutions of (\ref{eq:fuKS}).}
\label{sec:algorithm}

The goal of this section is to describe the algorithm for rigorous integration of (\ref{eq:fueq}).
Our algorithm  stems from the  \emph{method of self-consistent  bounds}, which was introduced in \cite{ZM} and later developed in \cite{ZAKS,ZGal,ZNS,ZKSper}. In the present paper we propose some significant modifications to the method.  The new  algorithm  combines the High Order Enclosure method \cite{NJP}  with the dissipative enclosure from \cite{ZKSper}. The other important novelty is the application of the automatic differentiation \cite{G,HNW} to the  infinite lader of ODEs represented by (\ref{eq:fueq}),  therefore the use of differential inclusions as in \cite{ZKSper} has been avoided.

Parts of our presentations will be abstract and directly applicable to system (\ref{eq:fueq}), while some others related to the automatic differentiation
will be tailored  to  (\ref{eq:fuKS}).

\subsection{The setting and abstract assumptions.}

We will write our system as
\begin{equation}\label{eq:generalODE}
 a'(t) =f(a(t))
\end{equation}
where $f\colon l_2 \supset \dom(f) \to l_2$. By $f_k$ we denote the $k^{\mathrm{th}}$ component of $f$.
We assume, that the vector field (\ref{eq:generalODE}) is of the form
\begin{equation*}
f_k(a) = -L_ka_k +  N_k(a)
\end{equation*}
and we make standing assumptions \textbf{C1--C5} on $f$ listed below. We assume that there are constants
\begin{equation*}
   q>1,\quad \Lpow > 1
\end{equation*}
such that
\begin{description}
 \item[C1:]there  exist constants  $L^*\geq L_*>0$, $K\geq 0$,  such that
    \begin{equation*}
      L_*k^\Lpow\leq L_k \leq L^*k^\Lpow, \quad k>K,
    \end{equation*}
 \item[C2:] there exists $r$ with
 $$r \leq p,$$
 such that
 for any $S>0$ there exists $D=D(S,q)$ such that
 $$|N_k(a)|\leq  Dk^rq^{-k} \quad \mbox{for}\quad k\geq 1, a\in  \WqS{q}{S},$$
 \item[C3:]
     for any $S>0$ $f_k\colon \WqS{q}{S}\to \mathbb{R}$ is continuous for $k=1,2,\ldots$,
 \item[C4:] for any $S>0$  and $i,k=1,2,\dots$  the function $\frac{\partial N_k}{\partial a_i}:\WqS{q}{S} \to \mathbb{R}$ is continuous and \\
  there exists $\llog=\llog(S,q)$, such that for all $k \in N_+$ there holds
   \begin{equation*}
      -L_k + \frac{1}{2}\sum_{i \in \mathbb{N}_+} \left|\sup_{a \in  \WqS{q}{S} }\frac{\partial N_k}{\partial a_i}(a) \right| +
      \frac{1}{2}\sum_{i \in \mathbb{N}_+} \left|\sup_{a \in  \WqS{q}{S} }\frac{\partial N_i}{\partial a_k}(a) \right|\leq \llog,
   \end{equation*}
\item[C5:] for any $S>0$, $k=1,2,\dots$  and any partial derivative operator $D=\frac{\partial^n }{\partial a_{j_1}\dots \partial a_{j_n}}$ the function $DN:\WqS{q}{S} \to \mathbb{R}$ is continuous.
\end{description}
In Section~\ref{sec:ks-estimates} we will show that conditions \textbf{C1--C5} are satisfied for the system (\ref{eq:fuKS}).

 The assumption \textbf{C2} contains two kinds of conditions; an estimate of $N_k$ on $\WqS{q}{S}$ and  the inequality $r<\Lpow$. The estimate is satisfied for any $N(u)=P(u,Du,\dots,D^m u)$, where $P$ is  a polynomial function with the constant $r \geq m$ depending on $m$ and the degree of $P$.  It might turn out, however, that $r \geq \Lpow$. This  happens for example for the viscous Burgers equation or the Navier-Stokes equations with periodic boundary condition in 2D or 3D. This will be avoided if instead of demanding that $|a_k| \leq S q^{-|k|}$ we will consider wider class of sets defined by $|a_k| \leq \frac{S}{q^{ |k|} \cdot |k|^t}$, where $t \in \mathbb{N}_+$ --- see \cite{ZKS3,ZNS} for more details.

 The constant $\llog$ in the assumption \textbf{C4} is an upper estimate for the logarithmic norm of $\pi_{\leq n} f $ on $\pi_{\leq n}E$ for all $n$  --- see \cite{ZAKS,ZNS,ZGal}) and Section~\ref{subsec:logN}. This estimate will be used to obtain a uniform bound for the Lipschitz constants of the flow induced by Galerkin projections, which will allow us for a nice convergence argument in the proof Theorem~\ref{thm:FOE-enclosure}.

The assumption \textbf{C5} is necessary in order to define the Taylor method for (\ref{eq:generalODE}).

\subsection{ An outline of the algorithm.}
 The algorithm uses the following data structure to store in a computer memory a class of subsets of $\ltwoq$
\begin{equation*}
  \begin{array}{lll}
  \textbf{type}&\textbf{GBound}& \\
  \{\\
    & m \geq0 &: \text{ a natural number},\\
    & \widetilde a \subset \mathbb R^m &: \text{ an interval vector},\\
    & S\geq 0 &: \text{ a real number},\\
    & q >1&: \text{ a real number}.\\
  \}
  \end{array}
\end{equation*}
An object $\mathbf{GBound}(m,\widetilde a,S,q)$ represents a set of real sequences $(a_k)_{k>0}$
with geometric-like decay of coefficients
\begin{equation*}
  a_k\in
  \begin{cases}
    \widetilde a_k, & 1\leq k \leq m\\
    [-S,S]\cdot q^{-k}, & k>m
  \end{cases}.
\end{equation*}
In the sequel we will use often the following decomposition of a set $E=\mathbf{GBound}(m,\widetilde a,S,q)$
\begin{equation*}
  E=X_E \oplus W_E, \quad   X_E=\pi_{\leq m} E , \quad W_E=\pi_{> m} E.
\end{equation*}

\begin{definition}
 Let $N \subset l_2$ and $h>0$. We say that
  the set  $E \subset l_2$ is a \emph{rough enclosure} for $N$ and a time $h$, if $N \subset E$ and for any $a_0 \in N$ any  solution of (\ref{eq:generalODE}) with initial condition $a(0)=a_0$ is defined for $t \in [0,h]$ and $a([0,h]) \subset E$.
\end{definition}
In other words $E$ gives a priori bounds for solutions starting from $N$ over a time interval $[0,h]$.

For a set of initial conditions $N=\mathbf{GBound}(m,\widetilde a_N,S_N,q)$ and a time step $h>0$ the algorithm
\begin{enumerate}
 \item computes the rough enclosure $E=\mathbf{GBound}(m,\widetilde a_E,S_E,q)$ for the set $N$ and the time step $h$,
 \item computes a tighter set $M=\mathbf{GBound}(m,\widetilde a_M,S_M,q)\subset E$ such that for $a\in N$ there holds $a(h)\in M$.
\end{enumerate}

 If Step 1 fails, the algorithm stops and we should restart the computation with refined data -- we can split the initial condition or shorten the time step.

\subsection{Basic existence and convergence theorems  justifying the correctness of the algorithm.}

We work under the standing assumptions \textbf{C1--C5}.

\begin{lemma}
\label{lem:f-cont-on-WqS}
For any $S>0$ the vector field $f$ is continuous on $\WqS{q}{S}$.
\end{lemma}
\textbf{Proof:}
Let us fix $a \in \WqS{q}{S}$. We will prove the continuity of $f$ at $a$. Let us fix $\epsilon>0$ and let $\widetilde a \in \WqS{q}{S}$.

We have from  conditions \textbf{C1} and \textbf{C2}
\begin{eqnarray*}
  \|f(a) - f(\widetilde a)\| &\leq& \|\pi_{\leq n}f(a) - \pi_{\leq n}f(\widetilde a)\| +  \|\pi_{> n}f(a)\| + \| \pi_{> n}f(\widetilde a)\| \leq \\
  &\leq &  \|\pi_{\leq n}f(a) - \pi_{\leq n}f(\widetilde a)\| + 2 \sum_{|k|>n} Dk^r q^{-k} + 2 \sum_{|k|>n}L^* k^\Lpow q^{-k} .
\end{eqnarray*}
Let $n$ be big enough to have
\begin{equation*}
  2 \sum_{|k|>n} Dk^r q^{-k} + 2 \sum_{|k|>n}L^* k^\Lpow q^{-k} < \epsilon/2.
\end{equation*}
From \textbf{C3} it follows that $\pi_{\leq n}f$ is continuous. Hence, there exists $\delta>0$, such that
$$\text{if}\quad \|a-\widetilde a\| \leq \delta\quad\text{then}\quad\|\pi_{\leq n}f(a) - \pi_{\leq n}f(\widetilde a)\| \leq \epsilon/2.$$ Gathering these estimates, we obtain
\begin{equation*}
   \text{if}\quad\|a-\widetilde a\|\leq \delta \quad\text{then}\quad \|f(a) - f(\widetilde a)\| \leq \epsilon.
\end{equation*}

\qed

\begin{lemma}
\label{lem:Gal-proj-error}
For $S>0$  there holds
\begin{eqnarray*}
  \lim_{n \to \infty} \sup_{a \in \WqS{q}{S}} \|f(a) - f(\pi_{\leq n}a)\|=0.
\end{eqnarray*}
\end{lemma}
\textbf{Proof:}
Let us fix $S>0$ and $\epsilon>0$. From Lemma~\ref{lem:f-cont-on-WqS} and the compactness of $\WqS{q}{S}$ it follows that $f$ is uniformly continuous on $\WqS{q}{S}$. Therefore, there exists $\delta>0$ such that
\begin{equation}\label{eq:uni-cont}
  \mbox{if} \quad \|a-\widetilde a\|\leq \delta, \quad \text{then} \quad \|f(a)-f(\widetilde a)\| \leq \epsilon.
\end{equation}

Let $n_0$ be large enough so that $\|\pi_{>n} \WqS{q}{S}\| < \delta$ for $n \geq n_0$. Hence, for $n \geq n_0$
if follows  that
\begin{equation*}
  \|a - \pi_{\leq n}a\|=\|\pi_{> n}a\| < \delta, \qquad \forall a \in \WqS{q}{S}.
\end{equation*}
Combining this with (\ref{eq:uni-cont}) we obtain
\begin{equation*}
  \|f(a) - f(\pi_{\leq n}a)\| < \epsilon, \quad \forall a \in \WqS{q}{S}.
\end{equation*}
This finishes the proof.
\qed

\begin{definition}
Let $N\subset \WqS{q}{S}$ be a set of initial conditions for (\ref{eq:generalODE}) and let us fix the time step $h>0$. We call the set $E=\mathbf{GBound}(m,\widetilde a_E,S_E,q)=X_E\oplus W_E $ \emph{a First Order Enclosure} (FOE) for $N$ over the time step $h$ if the following conditions are satisfied:
\begin{eqnarray}
 N&\subset& E,\label{eq:enclosureInclusion}\\
  \pi_{\leq m}N &+& [0,h](f_1,\ldots,f_m)(E)\subset \inte_{\leq m} X_E,\label{eq:finiteDimIsolation}\\
  a_k f_k(a)&<&0\quad \text{ for }\ a\in E, |a_k|=S_Eq^{-k}, k>m. \label{eq:tailIsolation}
 \end{eqnarray}
 \end{definition}

The next lemma shows that assumptions \textbf{C1--C3} guarantee that a FOE of the form $E=\mathbf{GBound}(m,\widetilde a_E,S_E,q)$ always exists for $N=\WqS{q}{S}$ provided the time step $h$ is small enough.  In  \cite{ZKS3} this was done for polynomial bounds of the form $W=\left\{a \ | \ |a_k| \leq \frac{S}{|k|^t}\right\}$.
Later, in Thorem~\ref{thm:FOE-enclosure}, we will prove that FOE is indeed a rough enclosure.
\begin{lemma}\label{lem:FOE-exists}
  If $N\subset \WqS{q}{S}$, then there exist $h>0$ such that the set
  $$E=\mathbf{GBound}(m,\widetilde a_E,S_E,q):=\WqS{q}{2S}$$ is a FOE for $N$ over the time step $h$ for (\ref{eq:generalODE}).
\end{lemma}
\textbf{Proof:}
Let us set $S_E=2S$ and $$E=\WqS{q}{2S}.$$
Let $D=D(E)$ be the constant from \textbf{C2}.
Take a natural number
$$m>\max\left\{K,\left(\frac{D}{L_*S_E}\right)^{\frac{1}{\Lpow-r}}\right\}.$$

The set $E$ can be seen as $\mathbf{GBound}(m,\tilde{a},S_E,q)=X_E\oplus W_E$, where
\begin{equation*}
  \tilde{a}_k=[-S_E q^{-k},S_E q^k], \quad k=1,\dots,m.
\end{equation*}
By the construction of $E$ we have $N\subset E$ and therefore (\ref{eq:enclosureInclusion}) is satisfied.

By \textbf{C1} and \textbf{C2}, for $k>m$ and $|a_k|=S_Eq^{-k}$ there holds
\begin{multline*}
 a_kf_k(a) = a_k(-L_ka_k + N_k(a)) \leq \\  -L_ka_k^2 + |a_kN_k(a)|
 \leq -L_*k^\Lpow S_E^2q^{-2k} + S_EDk^r q^{-2k} \leq \\
 S_Eq^{-2k}k^r\left(-L_*S_Ek^{\Lpow-r} + D\right) < \\
 S_Eq^{-2k}k^r\left(-L_*S_Em^{\Lpow-r} + D\right).
\end{multline*}
The constant $m$ is chosen so that $-L_*S_Em^{\Lpow-r} + D<0$ which implies $a_kf_k(a)<0$ for $k>m$ and $|a_k|=S_Eq^{-k}$. Therefore (\ref{eq:tailIsolation}) is satisfied on $E$.

There remains to show that (\ref{eq:finiteDimIsolation}) is satisfied for sufficiently small $h>0$. Since $E$ is compact, by \textbf{C3} the set $(f_1,\ldots,f_m)(E)$ is bounded. Therefore, we can find $h>0$ small enough such that $|hf_k(a)|< Sq^{-k}$ for $a\in E$ and $k=1,\ldots,m$. Then, for $a\in N$ and $k=1,\ldots, m$ we have
$$
|a_k + [0,h]f_k(E)| < Sq^{-k} + Sq^{-k} = S_Eq^{-k}
$$
and thus (\ref{eq:finiteDimIsolation}) holds true.
\qed

\begin{theorem}
\label{thm:FOE-enclosure}
If $N \subset \WqS{q}{S}$ and $E=\mathbf{GBound}(m,\widetilde a_E,S_E,q)=\WqS{q}{2S}$ is a FOE for $N$ over the time step $h$, then
for any $\widehat{a} \in N$ there exists $a:[0,h]\to E$, a  solution  of (\ref{eq:generalODE}), such that $a(0)=\widehat{a}$, i.e. $E$ is a rough enclosure for $N$ and the time step $h$. This solution is unique under requirement that $a([0,h]) \subset E$.

Moreover for any two solutions $\widehat a,\overline a:[0,h]\to E$ there holds
\begin{equation}
  \|\widehat a(t) - \overline a(t)\| \leq e^{\llog(2S,q)}\|\widehat a(0)-\overline a(0)\|, \quad t \in [0,h].  \label{eq:Lipconst-full-pde}
\end{equation}
\end{theorem}
\textbf{Proof:}
We will show that Galerkin projections of (\ref{eq:generalODE}) with initial condition in $N$ converge to the unique solution of the full system and we will also obtain (\ref{eq:Lipconst-full-pde}) in the process.

The $n^{\mathrm{th}}$ Galerkin projection of (\ref{eq:generalODE}) can be written as follows
\begin{equation}\label{eq:gal-proj}
  a'=\pi_{\leq n} f(a), \quad a(0) = \widehat{a} \in \pi_{\leq n}\ltwoq.
\end{equation}
Observe that by assumptions \textbf{C1, C3} and \textbf{C4} the right hand side of (\ref{eq:gal-proj}) is locally Lipschitz, hence the solution to (\ref{eq:gal-proj}) is unique. Let $\varphi_n(t,a)$ be a local flow induced by (\ref{eq:gal-proj}).

In the sequel we will assume that $n \geq m$.

\textbf{Step 1.} We will show that $\pi_{\leq n}E$ is a priori bound for solutions of (\ref{eq:gal-proj}) with initial conditions in $\pi_{\leq n} N$, i.e. $\varphi_n(t,a)$ is defined for $t \in [0,h]$ and $a \in \pi_{\leq n} N$ and
\begin{equation}\label{eq:apriori-gal}
  \varphi_n([0,h],\pi_{\leq n}N) \subset  \pi_{\leq n}E.
\end{equation}
Indeed, since $\pi_{\leq n} E \subset E$  and $E$ is a FOE it follows, that
\begin{eqnarray}
  \pi_{\leq m}N &+& [0,h](f_1,\ldots,f_m)(\pi_{\leq n}E)\subset \inte_{\leq m} E,\label{eq:fnfiniteDimIsolation}\\
  a_k f_k(a)&<&0\quad \text{ for }\ a\in \pi_{\leq n}E,\quad |a_k|=S_Eq^{-k}, n \geq  k>m. \label{eq:fntailIsolation}
 \end{eqnarray}
 From (\ref{eq:fntailIsolation}) it follows that while $\varphi_n(t,a) \in \pi_{\leq n}E$, then for $k =m+1,\ldots,n$ the trajectory cannot reach the part of the boundary $\partial_{\leq n}E$ defined by $|a_k|=S_Eq^{-k}$, because it is "repulsive" due to $\frac{d|a_k|}{dt} <0$ . The condition (\ref{eq:fnfiniteDimIsolation}) implies that the boundary $\partial_{\leq m} E$ in the $k^{\mathrm{th}}$ direction, $k=1,\ldots,m$, cannot be reached for $t \leq h$, because it is simply too far. This  completes the proof of (\ref{eq:apriori-gal}).

\textbf{Step 2.} We will show that for any $\widehat{a} \in N$ the function
$t \mapsto \varphi_n(t,\pi_{\leq n}\widehat{a})$ converges uniformly on $[0,h]$.

From Lemma~\ref{lem:Gal-proj-error} it follows that
\begin{equation*}
\delta_n=\sup_{a \in E} \|\pi_{\leq n}f(a) - \pi_{\leq n}f(\pi_{\leq n}a)\| \to 0, \quad n \to \infty.
\end{equation*}

Let $l=\llog(2S,q)$ be a constant from condition \textbf{C4}. Using  Theorem~\ref{thm:logn-formulas} (see Appendix \ref{subsec:logN}) and the Gershgorin Theorem it is easy to see that $l$ is an upper estimate for the logarithmic norm of $\pi_{\leq n} f $ on $\pi_{\leq n}E$ for all $n$.

Therefore we can estimate the difference between two Galerkin projections as follows. Let us take $n_1 > n$ and let $\widehat{a}$, $\overline{a} \in N$.

Observe that
$y(t)=\pi_{\leq n} \varphi_{n_1}(t,\pi_{\leq n_1} \overline{a})$ satisfies for $t \in [0,h]$
\begin{eqnarray}
  y'(t)&=&\pi_{\leq n}f\left(\varphi_{n_1}(t,\pi_{\leq n_1}\overline{a})\right) \nonumber \\
   &=&\pi_{\leq n}f(y(t)) + \pi_{\leq n}f\left(\varphi_{n_1}(t,\pi_{\leq n_1}\overline{a})\right) - \pi_{\leq n}f\left(\pi_{\leq n}\varphi_{n_1}(t,\pi_{\leq n_1}\overline{a})\right)  \nonumber \\
  &=&\pi_{\leq n}f(y(t)) + \delta(t), \quad |\delta(t)| \leq \delta_n.  \label{eq:err-full-gal}
\end{eqnarray}
Therefore, from Lemma~\ref{lem:estmLogN} (see Appendix \ref{subsec:logN}) we obtain for $t \in [0,h]$
\begin{eqnarray}
 \|\varphi_n(t,\pi_{\leq n}\widehat{a}) - \varphi_{n_1}(t,\pi_{\leq n}\overline{a})\| \leq \label{eq:gal-diff-two}\\
\nonumber   \|\varphi_n(t,\pi_{\leq n}\widehat{a}) - \pi_{\leq n}\varphi_{n_1}(t,\pi_{\leq n}\overline{a})\| + \|\pi_{> n}\varphi_{n_1}(t,\pi_{\leq n}\overline{a})\| \leq \\
\nonumber   e^{t l } \|\pi_{\leq n}\widehat{a} - \pi_{\leq n}\overline{a}\| + \delta_n \frac{e^{lt}-1}{l} + \left(\sum_{|k|> n} S_E^2 q^{-2k}\right)^{1/2}.
\end{eqnarray}
Now if we take $\widehat{a}=\overline{a}$, then
\begin{equation*}
  \|\varphi_n(t,\pi_{\leq n}\widehat{a}) - \varphi_{n_1}(t,\pi_{\leq n}\widehat{a})\| \leq \delta_n \frac{e^{lt}-1}{l} + \left(\sum_{|k|> n} S_E^2 q^{-2k}\right)^{1/2} \to 0,
\end{equation*}
when $n \to \infty$ uniformly for $(t,\widehat{a}) \in [0,h]\times N$.

Therefore we can define
\begin{equation*}
  \varphi(t,a)=\lim_{n \to \infty} \varphi_n(t,\pi_{\leq n}a).
\end{equation*}
Obviously $\varphi(t,a)$ is continuous on $[0,h] \times N $ as the limit of uniformly  convergent sequence of continuous functions.

\textbf{Step 3.} We will show, that $a(t)=\varphi(t,a)$ is a solution of (\ref{eq:generalODE}). For each $n_1 \geq n$ we have
\begin{equation}\label{eq:gal-calk}
  \pi_{\leq n} \varphi_{n_1}(t,\pi_{\leq n_1}a) = \pi_{\leq n}a + \int_0^t \pi_{\leq n} f( \varphi_{n_1}(s,\pi_{\leq n_1}a))ds.
\end{equation}
From Lemma~\ref{lem:f-cont-on-WqS} and compactness of $E$ it follows that $f$ is uniformly continuous on $E$. This combined with the uniform convergence of $\varphi_n$ implies that $f( \varphi_{n_1}(s,\pi_{\leq n_1}a))$ converges uniformly with $n_1 \to \infty$ to $f( \varphi(s,a))$ for $(s,a) \in [0,h]\times N$.  Therefore the integral on the rhs of (\ref{eq:gal-calk}) converges and we obtain for any $n$
\begin{equation*}
  \pi_{\leq n} \varphi(t,a) = \pi_{\leq n}a + \int_0^t \pi_{\leq n} f( \varphi(s,a))ds
\end{equation*}
which implies that
\begin{equation*}
  \varphi(t,a) = a + \int_0^t f( \varphi(s,a))ds.
\end{equation*}
By differentiation we see that $t \to \varphi(t,a)$ is a solution of (\ref{eq:generalODE}). Observe that when passing to the limit with $n,n_1 \to \infty$ in (\ref{eq:gal-diff-two}) we obtain that for $\widehat{a},\overline{a} \in N$ and $t\in [0,h]$ the condition (\ref{eq:Lipconst-full-pde}) is satisfied.

\textbf{Step 4.} It remains to show that if $a:[0,\tmax] \to E$ is a solution of (\ref{eq:generalODE}) with $a(0) \in N$, then for $t \in [0,\min(\tmax,h)]$ holds $a(t)=\varphi(t,a(0))$.

By decreasing $h$ if necessary without any loss of generality we can assume that $h=\tmax$. Observe, that for any $n$
the function $y(t)=\pi_{\leq n} a(t)$  satisfies (\ref{eq:err-full-gal}) for $t \in [0,h]$.
Therefore, from Lemma~\ref{lem:estmLogN} we obtain that for $t \in [0,h]$ there holds
\begin{eqnarray*}
 \|\varphi(t,a(0)) - a(t)\| \leq \\
  \|\varphi(t,a(0)) -\varphi_n(t,\pi_{\leq n}a(0))\| + \| \varphi_n(t,\pi_{\leq n}a(0)) - \pi_{\leq n}a(t)\| + \|\pi_{> n}a(t)\|\leq \\
\nonumber \|\varphi(t,a(0)) -\varphi_n(t,\pi_{\leq n}a(0))\| +    \delta_n \frac{e^{lt}-1}{l} + \left(\sum_{|k|> n} S_E^2 q^{-2k}\right)^{1/2}.
\end{eqnarray*}
Now passing to the limit $n \to \infty$ we obtain that $a(t)=\varphi(t,a(0))$ for $t \in [0,h]$.

\qed

\begin{rem}
Observe that (\ref{eq:tailIsolation}) used to obtain (\ref{eq:fntailIsolation}) was of fundamental importance in the above proof, as it provides us  uniform a priori bounds for high modes for all Galerkin projections. This is the isolation property mentioned in the Introduction.
\end{rem}

\begin{rem}
Theorem~\ref{thm:FOE-enclosure} implies that for fixed $q>1$ we have a family of continuous local semiflows defined on $\WqS{q}{S}$ and parameterized by $S \in \mathbb{R}_+$. We can consider a 'union' of these semiflows to obtain a local semiflow $\varphi$ on entire $\ltwoq$. However, from the above reasoning does not follow, that $\varphi$ is continuous with respect to $x$. This can be easily obtained from the classical existence results for KS equation recalled in Appendix~\ref{subapp:KS-existence}.
\end{rem}

\begin{theorem}
\label{thm:ltwoq-invsubspace}
For any $\bar{a} \in \ltwoq$ there exists $a: [0,\infty) \to \ltwoq$ a unique solution of (\ref{eq:generalODE}) with initial condition $a(0)=\bar{a}$.
The induced semiflow (semigroup) $\varphi$ is continuous with respect to $t$ and $\bar{a}$.
\end{theorem}

Let us emphasise, that our computer assisted proof does not depend neither on Theorem~\ref{thm:ltwoq-invsubspace} nor Theorem~\ref{thm:ks-semi-dyn-pde}. We just need a local continuous semiflow defined on $\WqS{q}{S}$, where $S>0$ is a computable constant, which contains finite number of sets generated by the computer program. These include selected subsets of certain Poincar\'e sections and enclosure of their forward trajectories unless they reach another Poincar\'e section.

\subsection{Algorithm, more details but still on the abstract level.}

\textbf{Notation.} In the sequel we will use the following notation.  For a smooth function $F\colon I\subset \mathbb R\to \mathbb R$ by $\tc F {} i(t)$ we denote the $i^{\mathrm{th}}$ Taylor coefficient of $F$ at $t$. We also write $F^{[i]}$ if $t=0\in I$. The same convention applies to vector valued functions $F\colon I\to {l_2}$.

In Section~\ref{subsec:ks-auto-diff} we will show that the components $a_k(t)$ of the solutions to the KS equation are smooth and we give an algorithm for computation of $\tc{a}{k}{i}$ by means of automatic differentiation.  It is important to emphasise that $\tc a k i$ depends on  $\tc a k j$ for $j=0,\dots,i-1$, hence it can be expressed in terms of $\tc{a}{k}{0}$, only. Here we assume this knowledge.

\subsubsection{High Order Enclosure in finite dimension.}

In \cite{NJP} the authors propose an efficient algorithm for computation of a priori bound on the set of trajectories over a time step. It is called the High Order Enclosure method as it relies on high order Taylor expansion of the solutions. In what follows we will recall the algorithm from \cite{CR,NJP}. In the next section we will show, how it can be adopted to the case of infinite dimensional dissipative systems.

Consider an initial value problem for a  finite dimensional ODE
\begin{equation}\label{eq:fdIVP}
x' = g(x),\quad x(0)\in X\subset U,
\end{equation}
where $g\in \mathcal C^{d+1}(U,\mathbb R^n)$, $U\subset \mathbb R^n$ is open and $d$ is a natural number. Let us fix $\widetilde h>0$ -- this is a trial time step for the numerical method used to integrate the system (\ref{eq:fdIVP}). In practice, $\widetilde h$ is generated by another algorithm --- a time step predictor. For the reasoning given here it is irrelevant, what this value come from.

\begin{theorem}{\cite[Theorem 3]{CR}}
Let $\varepsilon>0$ be a tolerance per time step and let us set $R=[-\varepsilon,\varepsilon]^n$. Let
\begin{equation*}
 P = \sum_{i=0}^d \tc {X}{} i[0,\widetilde h]^i
\end{equation*}
and define $E=P+R$.  If
 \begin{equation}
 \label{eq:hoeCondition}
 Z:= \tc E{} {d+1}[0,\widetilde h]^{d+1}\subset \inte R
 \end{equation}
 then for all $x(0)\in X$ the solution to (\ref{eq:fdIVP}) exists for all $t\in[0,\widetilde h]$ and $x([0,\widetilde h])\subset P+Z$.
\end{theorem}
The condition (\ref{eq:hoeCondition}) may fail due to three reasons. First and the most common case is when the inclusion (\ref{eq:hoeCondition}) is not satisfied. Given that $0\in\inte R$, we can always find $h<\widetilde h$ such that
 \begin{equation*}
 Z:=\tc E{} {d+1}[0, h]^{d+1}\subset \inte R
 \end{equation*}
and the set $P+Z$ is a high order enclosure for (\ref{eq:fdIVP}) with (usually slightly) decreased time step $h<\widetilde h$.

Second reason for failure is the situation, when the Taylor coefficient $\tc E{} {d+1}$ cannot be computed. This may happen for instance, when division by zero in interval arithmetics occurs or the set $E$ is out of the domain of $g$. In this case we have to decrease the trial step $\widetilde h$ and repeat the entire procedure (recomputation of $\tc {X}{} i$ required for $P$ is not necessary). If the number of such repetitions exceeds some specified maximal value, the algorithm stops and returns \textbf{Failure}. This does not mean that the trajectories do not exist  --- the algorithm simply could not validate requested condition.

Third, and less common situation is when we cannot compute Taylor $\tc X {} i$ used to define $P$. In this case we cannot proceed and the algorithm returns \textbf{Failure}. Some higher level decisions, such as changing the order, splitting the initial condition $X$ or just giving up have to be made.

\subsubsection{High Order Enclosure in infinite dimension.}\label{sec:hoe}

The construction of the rough enclosure as FOE given in the  proof of Lemma~\ref{lem:FOE-exists} was not used in our computer assisted proof, because  it produces too much overestimation and we also prefer to keep the dimension $m$ constant along the trajectory. Therefore, a bit more sophisticated routines mixing together a dissipative enclosure from \cite{ZKS3} and high order enclosure from \cite{CR,NJP} are necessary.

Let us consider now an infinite dimensional dissipative PDE of the form (\ref{eq:generalODE}). Let $N=\textbf{GBound}(m,\widetilde a, S, q)$ be a set of initial conditions for (\ref{eq:generalODE}).

Assume $\widetilde h>0$ is a trial time step given from prediction. In the algorithm given below we expect that the dynamics on modes $a_k$, for $k>m$ is highly dissipative. The construction of a priori bound consists of the following steps.
\begin{enumerate}
 \item Compute
\begin{equation*}
 P_E = \sum_{i=0}^d \pi_{\leq m}\tc {N}{} i[0,\widetilde h]^i.
\end{equation*}
The algorithm for computation of $\tc {N}{} i$ for the KS equation will be given in Section~\ref{subsec:ks-auto-diff}.
\item Set $R=[-\varepsilon,\varepsilon]^m$ and predict an enclosure on the main modes of the form $X_E= P_E+R$.
\item Set $S_E=S$ and $h=\widetilde h$.
\item Define the set $E=X_E\oplus W_E$, where
$$
  (W_E)_k = \begin{cases}
         0 & k\leq m \\
         [-S_E,S_E]q^{-k} & k>m
        \end{cases}
$$
\item Check if the vector field is pointing inwards $E_k$ for $k>m$. This is the same condition (\ref{eq:tailIsolation}) as in FOE. If not satisfied, then we slightly enlarge $S_E$ and go to step 4.
\item Check inclusion
 \begin{equation*}
 Z:=\pi_{\leq m}\tc E{} {d+1}[0, \widetilde h]^{d+1}\subset \inte R
 \end{equation*}
 If the above fails, then find $h<\widetilde h$ such that
 \begin{equation*}
 Z:=\pi_{\leq m}\tc E{} {d+1}[0, h]^{d+1}\subset \inte R
 \end{equation*}
 holds true.
\end{enumerate}
The above algorithm may fail by the same three reasons, as in the finite dimensional case. In addition, the loop in which we enlarge the constant $S_E$ may exceed specified maximal limit.

If the algorithm does not fail, the computed set $(P_E+Z)\oplus W_E$ is a priori bound for $a([0,h])$ for $a\in N$, as guaranteed by steps 5 and 6.

\subsubsection{Computation of tight enclosure for $a(h)$.}\label{sec:Taylor-method}

Once we have the rough enclosure for $a([0,h])$, next we want to compute  tight bounds for  $a(h)$.

Let $N=\mathbf{GBound}(m,\widetilde a_N,S_N,q)$ be a set of initial conditions for (\ref{eq:generalODE}). Fix $h>0$ and assume that $E=\mathbf{GBound}(m,\widetilde a_E,S_E,q)$ is a rough enclosure for $N$ over the time step $h$. We will show, how we can bound the set $\left\{a(h) : a\in N\right\}$.

Let us fix an order of the Taylor method $d>0$. On the main modes $k\leq m$ we will bound $a_k(h)$ by an explicit formula. For $a\in N$ by the Taylor theorem we have
\begin{equation*}
 a_k(h)\in \sum_{i=0}^d \tc a k i h^i + R_k,
\end{equation*}
where
\begin{equation*}
 R_k = \left\{\tc a k {d+1}[0,h]^{d+1}\, |\, a\in E\right\}.
\end{equation*}

The bound on the tail $k>m$ will be computed using infinite set of differential inequalities. Take
$D=D(E)$ from \textbf{C2}. By  assumptions \textbf{C1--C2} we have $$a_k'(t)\leq -L_ka_k + Dk^rq^{-k} = -L_ka_k + N_k^+,$$
where $N_k^+:=Dk^rq^{-k}$. Then, for $t\in[0,h]$ and $k>m$ there holds
\begin{multline*}
a_k(t)\leq \frac{N_k^+}{L_k} + \left(a_k(0)-\frac{N_k^+}{L_k}\right) e^{-L_kt} \leq\\
\frac{Dk^rq^{-k}}{L_*k^\Lpow} + \left(S_Eq^{-k}-\frac{Dk^rq^{-k}}{L_*k^\Lpow}\right) e^{-L_kt} = \\
\frac{D}{L_*k^{\Lpow-r}}q^{-k} + \left(S_Eq^{-k}-\frac{D}{L_*k^{\Lpow-r}}q^{-k}\right) e^{-L_kt} = \\
q^{-k}\left(\frac{D}{L_*k^{\Lpow-r}} + \left(S_E-\frac{D}{L_*k^{\Lpow-r}}\right) e^{-L_kt}\right) \leq \\
q^{-k}\left(\frac{D}{L_*(m+1)^{\Lpow-r}} + S_E e^{-L_*(m+1)^st}\right).
\end{multline*}
In a similar way we can bound $a_k(t)$ from below. To sum up, we proved the following
\begin{lemma}
Let $N=\mathbf{GBound}(m,\widetilde a_N,S_N,q)$ be a set of initial conditions for (\ref{eq:generalODE}). Fix $h>0$ and assume that $E=\mathbf{GBound}(m,\widetilde a_E,S_E,q)$ is a rough enclosure for $N$ over the time step $h$.

Then for $a\in N$ and $k>m$ there holds
\begin{equation*}
 |a_k(h)| \leq Sq^{-k},
\end{equation*}
where
\begin{equation*}
 S = \min\left\{S_E,\frac{D}{L_*(m+1)^{\Lpow-r}} + S_E e^{-L_*(m+1)^\Lpow h}\right\}.
\end{equation*}
\end{lemma}

We have shown, that given an initial condition represented as \textbf{GBound} with decay $q>1$ we can always find a rough enclosure for sufficiently small time step $h>0$ and compute a bound on the trajectories over the time step $h$ as \textbf{GBound} with the same geometric decay $q$.  We use higher order time-derivatives $\tc a k i$ which do not belong to $\ltwoq$  to bound finitely many leading modes $a_k(h)$ by an explicit formula.

\section{Estimates specific for the  KS equation.}\label{sec:ks-estimates}

In Section~\ref{sec:algorithm} we outlined an algorithm for computing rigorous enclosures on the set of trajectories in a class of vector fields satisfying assumptions \textbf{C1--C5}. We also assume that a routine that allows to compute Taylor coefficients $\tc a k i$  of solutions to (\ref{eq:generalODE}) is provided. In this section we show, that these requirements are satisfied for the Kuramoto-Sivashinsky equation (\ref{eq:fuKS}). Analogous requirements for sets of the form $W=\left\{|a_k| \leq \frac{C}{|k|^s}\right\}$ for a class of dissipative PDEs (including also KS equation) has been proved in \cite{ZNS}.

\subsection{Conditions C1--C5.}

The system (\ref{eq:fuKS}) splits into linear and nonlinear part as follows
\begin{eqnarray*}
 a_k'(t) &=& -L_ka_k + N_k(a),\\
 L_k &=& k^2(1-\nu k^2),\\
 N_k(a) &=& = - k\sum_{n=1}^{k-1}a_na_{k-n} + 2k\sum_{n=1}^\infty a_na_{n+k}.
\end{eqnarray*}

\textbf{C1:} It is easily satisfied as $L_k$ is a polynomial in $k$ of degree $s=4$.

\textbf{C2:} For $a\in \WqS{q}{S}$ we have
\begin{multline*}
 |N_k(a)| \leq k\sum_{n=1}^{k-1}Sq^{-n}Sq^{n-k} + 2k\sum_{n=1}^\infty Sq^{-n}Sq^{-n-k} =\\ k(k-1)S^2q^{-k} + 2kS^2q^{-k}(q^2-1)^{-1} \leq \\
 2k^2q^{-k}S^2\left(1 + (q^2-1)^{-1}\right).
\end{multline*}
Thus, the condition \textbf{C2} is satisfied with $r=2$ and $D =2S^2\left(1 + (q^2-1)^{-1}\right)$.

\textbf{C3:} We have to show that $f_k$ is continuous on $l_2$. The formula for $f_k$ splits into polynomial of degree $2$, which is clearly continuous, and an infinite sum
$$
a\to 2k\sum_{n=1}^\infty a_na_{n+k}
$$
which also is continuous as the inner product in $l_2$ composed with the $k$-shift of coordinates.

\textbf{C4:}
 Observe that
\begin{equation*}
  \frac{\partial N_k}{\partial a_{i}}= \left\{
                                         \begin{array}{ll}
                                           -2k a_{k-i} + 2k a_{k+i}, & \hbox{if $i <k$;} \\
                                           2k a_{k+i}, & \hbox{otherwise.}
                                         \end{array}
                                       \right.
\end{equation*}
 We estimate
\begin{equation*}
  \left|\frac{\partial N_k}{\partial u_{i}}(\WqS{q}{S})\right| \leq
  4 k S q^{-|k-i|}.
\end{equation*}
Hence
\begin{multline*}
 -L_k + \frac{1}{2}\sum_{i \in \mathbb{N}_+} \left|\sup_{a \in  \WqS{q}{S} }\frac{\partial N_k}{\partial a_i}(a) \right| +
      \frac{1}{2}\sum_{i \in \mathbb{N}_+} \left|\sup_{a \in  \WqS{q}{S} }\frac{\partial N_i}{\partial a_k}(a) \right| \leq \\
      -L_k  + 4kS\left(\sum_{i\in\mathbb{N}_+} q^{-|i-k|}\right) \leq \\
   -L_k  + 8k S \left( \sum_{i \in \mathbb{N}} q^{-i}\right) = -L_k +  \frac{8 k S}{1-q}.
\end{multline*}
Since  $L_k \approx \nu |k|^4$, we see that the above expression is bounded from above and $l_{\log}$ exists.

\textbf{C5:} It is obvious because $N$ is a quadratic polynomial.

\subsection{Coefficients $\tc{a}{k}{i}$ via automatic differentiation}
\label{subsec:ks-auto-diff}

Let us fix an initial condition $a(0)\in l_{2,q}$. In this section prove that the Taylor coefficients $\tc a{}{i}$ for the solutions to (\ref{eq:fuKS}) do exist. We propose an iterative scheme for computation bounds for $\tc a{}{i}$. These estimates will be also bounds for the Taylor coefficients for all Galerkin projections with  sufficiently high projection dimension. The proposed scheme is a special case of the general algorithm from \cite{WZ} tailored to the KS equation.

We will see that if $a(0) = \tc a{} 0\in \ltwoq$ then $\tc a {} i$, $i>0$  cannot belong to  $\ltwoq$. Therefore, we will construct a strictly decreasing sequence $q=q_0>q_1>\ldots >1$ and represent $\tc a {} i$ as geometric bounds in $l_{2,q_i}$.

The nonlinear part of (\ref{eq:fuKS})
\begin{equation*}
 N_k(t,a) = - k\sum_{n=1}^{k-1}a_n(t)a_{k-n}(t) + 2k\sum_{n=1}^\infty a_n(t)a_{n+k}(t)
\end{equation*}
splits into a finite sum
\begin{equation}\label{eq:explicitPart}
 E_k(t,a) = -\sum_{n=1}^{k-1}a_n(t)a_{k-n}(t) + 2\sum_{n=1}^m a_n(t)a_{n+k}(t)
\end{equation}
and an infinite sum
\begin{equation*}
 I_k(t,a) = \sum_{n=m+1}^\infty a_n(t)a_{n+k}(t).
\end{equation*}
Using this notation, the $k^{\textrm{th}}$ component of the vector field along the trajectory $a(t)$ reads
\begin{equation*}
F_k(t) = k\left(k(1-\nu k^2)a_k(t) + E_k(t,a(t)) + 2I_k(t,a(t))\right).
\end{equation*}
The coefficients $\tc a k i$ can be computed by the following
iterative scheme
\begin{equation} \label{eq:ak-taylor}
\tc a k {i+1} = \frac{1}{i+1}\tc F k i,
\end{equation}
provided the rhs of (\ref{eq:ak-taylor}) makes sense.

In what follows we will show, how we can bound $\tc E k i$,  $\tc I k i$ (where the Taylor coefficients are taken with respect to time variable) and, in consequence, to compute bounds on $\tc F k i$.

The estimates on the infinite part $\tc I k i$ are derived in the following
lemma.
\begin{lemma}\label{lem:DI}
Assume that for $j=0,1,\ldots,i-1$ we have already computed $\tc a {} i$ and
they are represented as geometric bounds $\tc a
{} i\in \mathbf{GBound}(m,\tc {\widetilde a} {} i,S_i,q_i)$. Then for $k\geq 1$ there holds that
\begin{equation*}
 \left|\tc I k i\right| \leq D_Iq^{-k},
\end{equation*}
where
\begin{eqnarray*}
 q&=& \min\{q_0,q_1,\ldots,q_i\},\\
 D_I&=&\left(\sum_{j=0}^iS_jS_{i-j}\right)\left((q^2-1)q^{2m} \right)^{-1}.
\end{eqnarray*}
\end{lemma}
\textbf{Proof:}
All the terms $a_n$ and $a_{n+k}$ which appear in the summation are bounded by
geometric series. Therefore we have
\begin{multline*}
\left| \tc I k i\right|
      \leq \sum_{n=m+1}^\infty\sum_{j=0}^i \left|\tc a n j \tc a {n+k} {i-j}\right|
      \leq \sum_{n=m+1}^\infty\sum_{j=0}^i S_jq_j^{-n}S_{i-j}q_{i-j}^{-n-k} \\
      \leq q^{-k}\left(\sum_{j=0}^i S_jS_{i-j}\right)\sum_{n=m+1}^\infty q^{-2n}\\
      = q^{-k}\left(\sum_{j=0}^iS_jS_{i-j}\right)\left((q^2-1)q^{2m}\right)^{-1}.
\end{multline*}
\qed

$E_k$ is handled by the following three lemmas.

\begin{lemma}\label{lem:Eleq2m}
For $k=1,\ldots,2m$ there holds that
\begin{multline*}
 \tc E k i = 2\sum_{j=0}^i \left(\sum_{n=1}^{\left\lfloor (k-1)/2\right\rfloor}
     \tc a n j \left(-\tc a {k-n} {i-j} + \tc a {n+k} {i-j}\right)
     +\sum_{n=\left\lceil k/2\right\rceil}^{m}\tc a n j \tc a {n+k} {i-j}\right)\\
     - \epsilon(k)\left(
     2 \sum_{j=0}^{\left\lfloor (i-1)/2\right\rfloor} \tc a {k/2} j \tc a {k/2} {i-j}
     +\epsilon(i) \left(\tc a {k/2} {i/2} \right)^2
     \right),
\end{multline*}
where $\epsilon(k)=1$ if $k$ is an even number and $\epsilon(k)=0$, otherwise.
\end{lemma}
\textbf{Proof:}
This is a direct application of the Leibniz rule to a slightly
factorized form of (\ref{eq:explicitPart}). This factorization is performed in order to reduce the computational cost and the wrapping effect in evaluation of this expression in interval arithmetics.
\qed


\begin{lemma}\label{lem:D1}
Assume that for $j=0,1,\ldots,i-1$ we have already computed $\tc a {} i$ and
they are represented as geometric bounds $\tc a {} i\in \mathbf{GBound}(m,\tc {\tilde a} {} i,S_i,q_i)$. Then for $k> 2m$ there holds that
\begin{equation*}
 \left|\tc E k i\right| \leq D_1 k q^{-k},
\end{equation*}
where
\begin{eqnarray*}
 q&=& \min\{q_0,q_1,\ldots,q_i\},\\
 D_1 &=& \frac{2}{2m+1}\sum_{j=0}^i\sum_{n=1}^m \left(q_{i-j}^n+q_{i-j}^{-n}\right) S_{i-j}\left|\tc a n j\right| + \sum_{j=0}^i S_jS_{i-j}.
\end{eqnarray*}
\end{lemma}
\textbf{Proof:}
For $k>2m$ the formula for $\tc E k i$ splits into two components
 $\tc E k i = \Sigma_1 + \Sigma_2$, where
\begin{eqnarray*}
 \Sigma_1 &=&  2\sum_{j=0}^i\sum_{n=1}^m \tc a n j \left(-\tc a {k-n} {i-j} + \tc a {n+k} {i-j}\right),\\
 \Sigma_2 &=& -\sum_{j=0}^i\sum_{n=m+1}^{k-1-m} \tc a n j \tc a {k-n} {i-j}.
\end{eqnarray*}
In $\Sigma_1$ the indices $k-n$ and $n+k$ are greater than $m$. Therefore
\begin{multline*}
|\Sigma_1|\leq
    2\sum_{n=1}^m \sum_{j=0}^i\left|\tc a n j\right| \left(S_{i-j}q_{i-j}^{n-k}+S_{i-j}q_{i-j}^{-n-k}\right)\\
    \leq q^{-k}\left(
        2\sum_{n=1}^m \sum_{j=0}^i\left(q_{i-j}^n+q_{i-j}^{-n}\right) S_{i-j}\left|\tc a n j\right|
    \right)\\
    \leq kq^{-k}\left(
        \frac{2}{2m+1}\sum_{n=1}^m \sum_{j=0}^i\left(q_{i-j}^n+q_{i-j}^{-n}\right) S_{i-j}\left|\tc a n j\right|
    \right).
\end{multline*}
In the sum $\Sigma_2$ we have $n>m$ and $k-n>m$. Therefore
\begin{equation*}
|\Sigma_2|\leq
    \sum_{n=m+1}^{k-1-m} \sum_{j=0}^iS_jq_j^{-n} S_{i-j}q_{i-j}^{n-k}.
\end{equation*}
For $k>n$ and $i=0,\ldots,j$ there holds that
$$
q_j^{-n}q_{i-j}^{n-k} = q^{-k}\left(\frac{q}{q_j}\right)^n\left(\frac{q}{q_{i-j}}\right)^{k-n}\leq q^{-k}.
$$
This gives us an estimate
\begin{equation*}
|\Sigma_2|\leq q^{-k}\sum_{n=m+1}^{k-1-m}\sum_{j=0}^i S_jS_{i-j}
 \leq kq^{-k}\left(\sum_{j=0}^i S_jS_{i-j}\right)
\end{equation*}
Gathering together bounds on $\Sigma_1,\Sigma_2$ we obtain the constant $D_1$.
\qed


\begin{lemma}\label{lem:D2}
Assume that for $j=0,1,\ldots,i-1$ we have already computed $\tc a {} i$ and
they are represented as geometric bounds $\tc a {} i\in \mathbf{GBound}(m,\tc {\tilde a} {} i,S_i,q_i)$.
Then for $m<k\leq 2m$ there holds that
\begin{equation*}
 |\tc E k i| \leq D_2 k q^{-k},
\end{equation*}
where
\begin{eqnarray*}
 q&=& \min\{q_0,q_1,\ldots,q_i\},\\
 D_2 &=& \max_{k=m+1,\ldots,2m}\frac{1}{k}\left|\tc E k i\right|q^k.
\end{eqnarray*}
\end{lemma}
\textbf{Proof:}
$\tc E k i$ is given by an explicit finite sum. Thus we can bound $D_2$ in finite computation.
\qed

All the above considerations lead to the following estimate on the $i^{\mathrm{th}}$ Taylor coefficient of $F$.
\begin{lemma}\label{lem:updateQi}
Let $q=\min\{q_0,\ldots,q_i\}$ and fix $\delta\in(1,q)$. Put
\begin{equation*}
L=\begin{cases}
    (m+1)^4\delta^{-(m+1)}& \text{if } m>4/\ln\delta,\\
    \left(\frac{4}{e\ln\delta}\right)^4& \text{otherwise}.
\end{cases}
\end{equation*}
Let $D_I, D_1, D_2$ be constants computed as in Lemma~\ref{lem:DI}, Lemma~\ref{lem:D1} and Lemma~\ref{lem:D2}, respectively,  and put $D=\max\{D_1,D_2\}$. Then for $k>m$ there holds that
\begin{equation*}
\left| \tc F k i \right | \leq S (q/\delta)^{-k},
\end{equation*}
where
\begin{equation*}
 S = L\left(|\nu- (m+1)^{-2}| S_i + (m+1)^{-3}D_I+ (m+1)^{-2}D\right).
\end{equation*}
\end{lemma}
\textbf{Proof:}
$\tc F k i$ is given by
$$
\tc F k i = k^2(1-\nu k^2)\tc a k i + k(\tc E k i + 2 \tc I k i).
$$
For $k>m$ we have
\begin{eqnarray*}
\left|\tc F k i\right| &\leq& k^2|1-\nu k^2| S_iq_i^{-k} + kD_Iq^{-k} + k^2Dq^{-k}\\
&\leq& \left(k^2|1-\nu k^2| S_i + kD_I + k^2D\right)q^{-k}\\
&=& \left(|\nu- k^{-2}| S_i + k^{-3}D_I + k^{-2}D\right)k^4q^{-k}\\
&\leq& \left(|\nu- (m+1)^{-2}| S_i + (m+1)^{-3}D_I+ (m+1)^{-2}D\right)k^4q^{-k}.
\end{eqnarray*}
We would like to find a constant $L$ such that
$$k^4q^{-k}\leq L(q/\delta)^{-k}$$
for $k>m$. Thus $L$ must satisfy $L\geq k^4\delta^{-k}$ for $k>m$. The function $k\to k^4\delta^{-k}$ attains the global maximum at $k=4/\ln \delta$ and it is strictly decreasing to zero after reaching maximum at this point. If $m>4/\ln\delta$ then $L$ can be taken as $L=(m+1)^4\delta^{-m-1}$. Otherwise we have to take the global maximum $L=\left(\frac{4}{e\ln\delta}\right)^4$.
\qed

Observe that the above derivations are valid also for all $n$-dimensional  Galerkin projections with $n \geq m$. Indeed in this situation we
have $$\pi_{\leq n}\mathbf{GBound}(m,{\tilde a},S_i,q_i) \subset \mathbf{GBound}(m,{\tilde a},S_i,q_i).$$

From the above Lemmas we immediately obtain the following
\begin{rem}
Let $a:[0,h] \to l_{2,q}$ be a solution of the KS equation (or its $n$-dimensional Galerkin projection with $n \geq m$), such that for $t\in[0,h]$ there holds
$a(t) \in \mathbf{GBound}(m,{\tilde a},S_0,q_0=q)$. Then for any $i \in \mathbb{N}$ there exists $\tc{a}{k}{i}(t)$ for $t \in
[0,h]$. Moreover,  it satisfies
\begin{equation*}
  |\tc{a}{k}{i}(t)| \leq \frac{S(q/\delta)^{-k}}{i+1}, \quad k >m
\end{equation*}
where $S$ and $\delta$ are as in Lemma~\ref{lem:updateQi}.
The formulas for $k \leq m$ follow from Lemmas~\ref{lem:DI} and~\ref{lem:Eleq2m}.
\end{rem} 

\appendix

\section{Appendix}

\subsection{The existence and uniqueness results for Kuramoto-Sivashinsky PDE.}
\label{subapp:KS-existence}

We recall results from Section III.4.1 in \cite{Temam}. There KS equation is written as
\begin{equation}\label{eq:T-Ks}
  \frac{\partial v}{\partial t} + \nu \frac{\partial^4 v}{\partial x^4} + \frac{\partial^2 v}{\partial x^2}+ v\frac{\partial v}{\partial x}=0
\end{equation}
on $\Omega=[-L/2,L/2]$ subject to periodic boundary conditions
\begin{equation}\label{eq:T-ks-bd}
  \frac{\partial^j v}{\partial x^j}\left(t,-\frac{L}{2}\right)= \frac{\partial^j v}{\partial x^j}\left(t,\frac{L}{2}\right), \quad j=0,1,2,3.
\end{equation}

Let
\begin{eqnarray*}
 H&=&\dot{L}(\Omega)=\left\{v \in L^2\left(\Omega\right)\, \left| \, \int_{-L/2}^{L/2} v(s)ds =0  \right. \right\},\\
 V&=& H^2_{per}(\Omega) \cap H.
\end{eqnarray*}
$H$ is endowed with $L^2$ scalar product  (denoted $(\cdot,\cdot)$), while $V$ is endowed with the scalar product
\begin{eqnarray*}
  ((v,w)) &=& \int_{-L/2}^{L/2} \frac{\partial^2 v}{\partial x^2}  \frac{\partial^2 w}{\partial x^2} dx.
\end{eqnarray*}
Let $A$ be an unbounded self-adjoint operator in $H$ with the domain $D(A)=H^4_{per} \cap H$ given by
\begin{equation*}
  Av=\left(\frac{\partial }{\partial x}\right)^4 v.
\end{equation*}

\begin{theorem}
\label{thm:ks-semi-dyn-pde}
For $v_0 \in H$ there exists a unique solution $v$ of (\ref{eq:T-Ks},\ref{eq:T-ks-bd}) with $v(0)=v_0$,
\begin{equation*}
  v \in \mathcal{C}([0,T]; H) \cap L^2(0,T;V),\quad \forall T>0.
\end{equation*}
Furthermore, for $t>0$ the function $v$ is analytic in $t$ with values in $D(A)$ and the mapping
\begin{equation*}
  v_0 \mapsto v(t)
\end{equation*}
is continuous from $H$ into $D(A)$.
\end{theorem}



\subsection{Logarithmic norms and estimates for Lipschitz constant of the flow induced by ODEs.}
\label{subsec:logN}
The goal is to recall some results  about the Lipschitz constant  of the flow induced by ODEs based on the logarithmic norms.

\begin{definition} \cite[Def. I.10.4]{HNW}
Let $Q$ be a square matrix;  we call
\begin{displaymath}
  \mu(Q) = \lim_{h > 0 ,h\to 0} \frac{\|I + hQ \| - 1}{h}
\end{displaymath}
the {\em logarithmic norm} of $Q$.
\end{definition}

\begin{theorem} \cite[Th. I.10.5]{HNW}
\label{thm:logn-formulas}
The logarithmic norm is obtained by the following formulas
\begin{itemize}
\item for Euclidean norm
  \begin{displaymath}
     \mu(Q)=\text{the largest eigenvalue of}\ 1/2(Q + Q^T),
  \end{displaymath}
\item for max norm $\|x\|_\infty=\max_{k} |x_k|$
  \begin{displaymath}
     \mu(Q)=\max_k \left( q_{kk} + \sum_{i\neq k} |q_{ki}| \right),
  \end{displaymath}
\item for norm $\|x\|_1=\sum_k  |x_k|$
  \begin{displaymath}
     \mu(Q)=\max_i \left( q_{ii} + \sum_{k\neq i} |q_{ki}| \right).
  \end{displaymath}
\end{itemize}
\end{theorem}

Consider now the differential equation
\begin{equation}
  x'=f(x), \quad \mbox{$f \in \mathcal C^1(\mathbb R^n)$}. \label{eq:odelogn}
\end{equation}
Let $\varphi(t,x_0)$  denote the solution of equation
(\ref{eq:odelogn}) with the initial condition $x(0)=x_0$. By $\|x
\|$ we denote a fixed arbitrary norm in $\mathbb{R}^n$.

The following theorem was proved in \cite[Th. I.10.6]{HNW} (for a
non-autonoumous ODE, here we restrict ourselves  to the autonomous
case only and we use a different notation).
\begin{theorem}
Let $y:[0,T] \to \mathbb{R}^n$ be a piecewise
$\mathcal C^1$ function and $\varphi(\cdot,x_0)$ be defined for $t \in
[0,T]$. Suppose that  the following estimates hold:
\begin{eqnarray*}
  \mu\left(\frac{\partial f}{\partial x}(\eta)\right) \leq l(t),\quad \mbox{ for $\eta \in [y(t),\varphi(t,x_0)] $}, \\
  \left\| \frac{dy}{dt}(t) - f(y(t)) \right\| \leq \delta(t),
\end{eqnarray*}
where $[y(t),\varphi(t,x_0)]$ denotes the line segment  connecting $y(t)$ and $\varphi(t,x_0)$.

Then for $0 \leq t \leq T$ there holds
\begin{displaymath}
 \| \varphi(t,x_0) - y(t)  \| \leq e^{L(t)}\left( \|y(0) - x_0 \| + \int_0^te^{-L(s)}\delta(s)ds  \right),
\end{displaymath}
where $L(t)=\int_0^tl(s)ds$.
\end{theorem}

From the above theorem one easily derives the following.
\begin{lemma}
\label{lem:estmLogN} Let $y:[0,T] \to \mathbb{R}^n$ be a piecewise
$\mathcal C^1$ function and $\varphi(\cdot,x_0)$ be defined for $t \in
[0,T]$. Suppose that $Z$ is a convex set such that   the following
estimates hold:
\begin{eqnarray*}
  y([0,T]), \varphi([0,T], x_0) \in Z,  \\
  \mu\left(\frac{\partial f}{\partial x}(\eta)\right) \leq l,\quad \mbox{ for $\eta \in Z$}, \\
  \left\| \frac{dy}{dt}(t) - f(y(t)) \right\| \leq \delta.
\end{eqnarray*}
Then for $0 \leq t \leq T$ there holds
\begin{displaymath}
 \| \varphi(t,x_0) - y(t)  \| \leq e^{lt}  \|y(0) - x_0 \| + \delta \frac{e^{lt} -1}{l},\quad \mbox{if $l \neq 0$}.
\end{displaymath}
For $l=0$, there holds
\begin{displaymath}
 \| \varphi(t,x_0) - y(t)  \| \leq  \|y(0) - x_0 \| + \delta t.
\end{displaymath}
\end{lemma}

\subsection{The Poincar\'e transition maps between sections.}
\label{subapp:poincarePDE}

Let $\Pi_1$ and $\Pi_2$ be affine hyperplanes in $l_2$ given by $v_i(x)+c_i=0$ for $i=1,2$, where $v_i$ the continuous linear forms.

In order to define the Poincar\'e transition map between $\Pi_1$ and $\Pi_2$ for the local semiflow $\varphi(t,x)$ induced by (\ref{eq:generalODE})  we need the following
\begin{itemize}
  \item continuity of $\varphi(t,x)$ with respect to both variables,
  \item transversality of the trajectories $t \mapsto \varphi(t,x)$, when intersecting $\Pi_i$. For the ODE case it is enough to have $v_i(f(y)) \neq 0$ for $y$ in the neighborhood of  intersection of $\varphi(t,x)$ with section $\Pi_i$.
\end{itemize}
In our case we will demand that analogous conditions hold in some $\WqS{q}{S}$ close to the starting and the target sections.

We assume the standing assumptions \textbf{C1--C5} and that there exists $m$, such that all forms $v_i$ are defined in terms of first $m$ variables, i.e. $v_i(a)=v_i(\pi_{\leq m}a)$.

\begin{definition}
We define \emph{a global section} as a hyperplane:
\begin{equation*}
\Pi = \{x \in l_2 \,  | \,  v(x) + c = 0\},
\end{equation*}
where $v:l_2 \to \mathbb{R}$, such that $v(a)=v(\pi_{\leq m}a)$ and $c \in \mathbb{R}$.

Any convex and bounded subset $U \subset \Pi$ is called
\emph{a local section}.
\end{definition}

Let $W_Z={\widetilde a} \oplus \pi_{>m} \WqS{q}{S}$, where ${\widetilde a}$ is an open convex subset of $\pi_{\leq m} l_2$.
\begin{definition}
A local section $Z$ is said to be \emph{transversal} in $W_Z$ if
\begin{equation*}
  W_Z \cap Z = Z, \qquad W_Z=W_{Z,-} \cup Z \cup W_{Z,+},
\end{equation*}
where
\begin{equation*}
  W_{Z,-} = \{x \in W_Z \ | \  v(x)+c<0\}, \qquad W_{Z,+}=\{x \in W_Z \ | \ v(x)+c>0\},
\end{equation*}
satisfying the condition
\begin{equation}
\label{eq:sec-trans}
  v\left( f(x)) \right) > 0, \quad \forall x \in W_Z.
\end{equation}
We will refer to (\ref{eq:sec-trans}) as the \emph{transversality condition}.
\end{definition}

We have the following easy lemma.

\begin{lemma}
Let $Z$ be a local transversal section in $W_{Z}$ for (\ref{eq:generalODE}) and let $N \subset \WqS{q}{S}$ for some $S>0$.
Assume that there exist $t_1,t_2 \in \mathbb{R}$, $ t_1 < t_2$, such that the following
conditions hold for all $x \in N$:
\begin{eqnarray*}
\varphi((t_1,t_2),x) \subset W_Z, \quad \varphi(t_1,x) \in W_{Z,-} & \quad and \quad & \varphi(t_2,x) \in W_{Z,+} .
\end{eqnarray*}

Then, for each $x_0 \in N$, there exists unique $t_{Z}(x_0) \in (t_1, t_2)$
such that $\varphi\left(t_{Z}(x_0), x_0\right) \in Z$.
Also, $t_{Z} : N \to [t_1, t_2]$ is continuous.
\end{lemma}

Using the above lemma we can define a map $P_{Z_1 \to Z_2}:Z_1 \to Z_2$ for two transversal local sections $Z_1$ and $Z_2$, by
$$P_{Z_1 \to Z_2}(x)=\varphi(t_{Z_2}(x),x).$$

\subsection{Representation of h-sets and computer-assisted verification of covering relations with one exit direction.}
\label{subapp:covrel-check}

Let us fix $q>1$.

In order to treat the system (\ref{eq:fuKS}) rigorously on a computer we define a data structure which represents $h$-sets in $\ltwoq$:

\begin{equation}\label{eq:hset-representation}
  \begin{array}{lll}
  \textbf{type}&\textbf{HSet}& \\
  \{\\
    & m \geq0 &: \text{ a natural number},\\
    & S\geq0 &: \text{ a real number},\\
    & x\in \mathbb R^m &: \text{ a vector},\\
    & A \in \mathbb R^{m^2} &: \text{ an invertible matrix}.\\
  \}
  \end{array}
\end{equation}
An $h$-set $N=\mathbf{HSet}(m_N,S_N,x_N,A_N)$ represented by the above data structure is given by
\begin{equation*}
 N = c_N\left([-1,1]^{m_N}\oplus W\right),
\end{equation*}
where $c_N$ is an invertible affine map
\begin{equation*}
 c_N\left((a_k)_{k=1}^\infty\right) = \left(x_N+A_N\cdot(a_1,\ldots,a_{m_N})^T,a_{m_N+1},a_{m_N+2},\ldots\right)
\end{equation*}
and
\begin{equation*}
 W = [-1,1]\cdot \left(\underbrace{0,0,\ldots,0}_{m_N},S_N q^{-(m_N+1)},S_N q^{-(m_N+2)},\ldots\right).
\end{equation*}
Thus, the projection of $N$ onto $m$ leading coordinates is a parallelepiped centered at $x_N$ with the shape matrix $A_N$. The remaining coordinates are not affected by $c_N$ and are stored as a geometrically decaying tail.

We also make an assumption that all $h$-sets have one  exit (nominally unstable) direction, i.e. $u(N)=1$ in Definition~\ref{def:hset}. Thus, the tail of $N$ is always given by
$$
\overline{T_N} = [-1,1]\cdot\left(0,\underbrace{1,1,\ldots,1}_{m_N-1},S_N q^{-(m_N+1)},S_N q^{-(m_N+2)},\ldots\right).
$$

Let $N_0$ and $N_1$ be $h$-sets contained in $\ltwoq$ and represented as in (\ref{eq:hset-representation}). Let $f:|N_0|\to l_2$ be continuous and compact map and such that $f(|N_0|) \subset \WqS{q}{S}$, for some $S>0$.

 Assume that we have a routine which for given $A\subset |N|$ computes $B \subset \WqS{q}{S_1}$, also represented by (\ref{eq:hset-representation}), such that  $\left(c_{N_1}^{-1}\circ f\circ c_{N_0}\right)(A)\subset B$.

Put
\begin{equation}\label{eq:conditionsForCoveringRelations}
\begin{array}{rcl}
B &=& \left(c_{N_1}^{-1}\circ f\circ c_{N_0}\right)([-1,1]\oplus \overline{T_{N_0}}),\\
B_l &=& \left(c_{N_1}^{-1}\circ f\circ c_{N_0}\right)(\{-1\}\oplus \overline{T_{N_0}}),\\
B_r &=& \left(c_{N_1}^{-1}\circ f\circ c_{N_0}\right)(\{1\}\oplus \overline{T_{N_0}}).
\end{array}
\end{equation}
Now,  checking $N_0\cover{f}N_1$ reduces to a finite set of inequalities which must be satisfied. These are:
\begin{itemize}
 \item $\pi_{i}(B)\subset (-1,1)$ for $i=2,\ldots, m$,
 \item $S(B)<S(N_1)$, where $S$ is a positive constant in $h$-set representation (\ref{eq:hset-representation}),
 \item either $\pi_1(B_l)<-1$ and $\pi_1(B_r)>1$ or $\pi_1(B_l)>1$ and $\pi_1(B_r)<-1$.
\end{itemize}
The two cases in the last condition depend on whether the mapping $f$ changes or not the orientation along the exit  direction.

It is easy to see that
\begin{eqnarray*}
H(t,\cdot) &=& (1-t) \left(c_{N_1}^{-1}\circ f\circ c_{N_0}\right) + tL,\\
L((a_k)_{k=1}^\infty) &=& (\pm 2a_1,0,0\ldots),
\end{eqnarray*}
are a homotopy and a linear map required by Definition~\ref{def:covrel}, where the sign in $L$ depends on whether $f$ preserves or not the orientation in the exit direction.

\subsection{Technical data.}\label{sec:technical-data}
The source code of the C++11 program that realises the computer-assisted proof of Lemma~\ref{lem:symbolic-dynamics} is available at \cite{W}. Below we list our choices of some parameters of the algorithms.
\begin{itemize}
 \item All $h$-sets that appear in Lemma~\ref{lem:symbolic-dynamics} are represented as a data structure (\ref{eq:hset-representation}) with the constant $m=15$.
 \item We set $d=4$ as the order of the Taylor method (Section~\ref{sec:Taylor-method}) for rigorous integration of (\ref{eq:fuKS}).
 \item High-Order Enclosure (Section~\ref{sec:hoe}) with $d=4$ acts on $m=11$ number of modes.
\end{itemize}
Verification of covering relation $N_0\cover{f}{N_1}$ requires computation of three images of $f$ --- see (\ref{eq:conditionsForCoveringRelations}). In Lemma~\ref{lem:symbolic-dynamics} we verify $26$ covering relations which means, that we have to verify $78=3\cdot 26$ inequalities. Due to unavoidable overestimation in computation of Poincar\'e maps we had to subdivide few initial conditions. Thus, the total number of images of Poincar\'e map we had to compute was $84$.

We run the program which checks all the inequalities required for covering relations listed in Lemma~\ref{lem:symbolic-dynamics} on a computer equipped with $120$ Intel(R) Xeon(R) CPU E7-8867 v3 @ 2.50GHz processors ($84$ of them were used). The program completed verification within $86$ minutes, which is the CPU time needed for the longest integration in $N_{2\to 1}^8\cover{P^6}N_{2\to1}^9$.

The algorithm for rigorous integration of the KS equation is a part of the CAPD library \cite{CAPD}. We tested the program with CAPD ver. 5.0.34 and the C++11 compiler from gcc-5.2 suite.

\end{document}